\documentclass[10pt,twocolumn]{article}

\usepackage{fullpage}             
\usepackage{graphicx}             
\usepackage{subfigure}            
\usepackage{url}                  
\usepackage[latin1]{inputenc}     
\usepackage{nicefrac}             
\usepackage[super,negative]{nth}  
\usepackage{indentfirst}          
\usepackage{algorithm}            
\usepackage{algpseudocode}        
\usepackage{amssymb}              

\newcommand{\dfn}[1]{\textit{#1}}            

\title{Efficient Algorithms for Large-Scale Topology Discovery}

\author{Benoit Donnet, Philippe Raoult, Timur Friedman, Mark
Crovella\thanks{Mr. Donnet, Mr. Raoult, and Mr. Friedman are with
the Laboratoire LiP6-CNRS of the Université Pierre et Marie Curie,
Paris.  Mr. Crovella is with the Computer Science Department,
Boston University. The authors are participants in the
traceroute@home project. This work was supported by: the RNRT
project Metropolis, NSF grants ANI-9986397 and CCR-0325701, a
SATIN European Doctoral Research Foundation grant, the e-Next
European Network of Excellence, and LiP6 2004 project funds.  This
work was performed while Mr. Crovella was at LiP6, with support
from the CNRS and Sprint Labs.}}

\date{}   

\begin{document}

\maketitle

\begin{abstract}
There is a growing interest in discovery of internet topology at
the interface level.  A new generation of highly distributed
measurement systems is currently being deployed. Unfortunately,
the research community has not examined the problem of how to
perform such measurements efficiently and in a network-friendly
manner.  In this paper we make two contributions toward that end.
First, we show that standard topology discovery methods (e.g.,
skitter) are quite inefficient, repeatedly probing the same
interfaces. This is a concern, because when scaled up, such
methods will generate so much traffic that they will begin to
resemble DDoS attacks. We measure two kinds of redundancy in
probing (intra- and inter-monitor) and show that both kinds are
important.  We show that straightforward approaches to addressing
these two kinds of redundancy must take opposite tacks, and are
thus fundamentally in conflict.  Our second contribution is to
propose and evaluate Doubletree, an algorithm that reduces both
types of redundancy simultaneously on routers and end systems. The
key ideas are to exploit the tree-like structure of routes to and
from a single point in order to guide when to stop probing, and to
probe each path by starting near its midpoint. Our results show
that Doubletree can reduce both types of measurement load on the
network dramatically, while permitting discovery of nearly the
same set of nodes and links.  We then show how to enable efficient
communication between monitors through the use of Bloom
filters.
\end{abstract}

\section*{Introduction}\label{introduction}

Systems for active measurements in the internet are undergoing a
radical shift.  Whereas the present generation of systems operates
on largely dedicated hosts, numbering between 20 and 200, a new
generation of easily downloadable measurement software means that
infrastructures based on thousands of hosts could spring up
literally overnight.  Unless carefully controlled, these new
systems have the potential to impose a heavy load on parts of the
network that are being measured.  They also have the potential to
raise alarms, as their traffic can easily resemble a distributed
denial of service (DDoS) attack.  This paper examines the problem,
and proposes and evaluates an algorithm for controlling one of the
most common forms of active measurement:
\dfn{traceroute}~\cite{traceroute}.

There are a number of systems active today that aim to elicit the
internet topology at the IP interface level. The most extensive
tracing system, \textsc{Caida}'s \dfn{skitter}~\cite{skitter},
uses 24 monitors, each targeting on the order of one million
destinations. Some other well known systems, such as the
\textsc{Ripe} NCC's \dfn{TTM service}~\cite{ripeNccTtm} and the
\textsc{NLanr} \dfn{AMP}~\cite{nlanrAmp}, have larger numbers of
monitors (between one- and two-hundred), and conduct traces in a
full mesh, but avoid tracing to outside destinations.

The uses of the raw data from these traces are numerous.  From a
scientific point of view, the results underlie efforts to model
the network~\cite{agarwal, connectivity, faloutsos, relationship,
asanalysis, assize}. From an  engineering standpoint, the results
inform a wide variety of protocol development choices, such as
multicast and overlay construction \cite{discovering}.

However, recent studies have shown that reliance upon a relatively
small number of monitors can introduce unwanted biases.  For
instance, work by Faloutsos et al.~\cite{faloutsos} found that the
distribution of router degrees follows a power law. That work was
based upon an internet topology collected from just twelve
traceroute hosts by Pansiot and Grad~\cite{onRoutes}.  However,
Lakhina et al.~\cite{sampling} showed that, in simulations of a
network in which the degree distribution does not at all follow a
power law, traceroutes conducted from a small number of monitors
can tend to induce a subgraph in which the node degree
distribution does follow a power law. Clauset and
Moore~\cite{tracerouteSampling} have since demonstrated
analytically that such a phenomenon is to be expected for the
specific case of the Erd\"os-R\'enyi random
graphs~\cite{erdosRenyi}.

Removing potential bias is not the only reason to employ
measurement systems that use a larger number of monitors. With
more monitors to probe the same space, each one can take a smaller
portion and probe it more frequently. Network dynamics that might
be missed by smaller systems can more readily be captured by the
larger ones.

The idea of releasing easily deployable measurement software is
not new.  To the best of our knowledge, the idea of incorporating
a traceroute monitor into a screen saver was first discussed in a
paper by Cheswick et al.~\cite{mapping} from the year 2000 (they
attribute the suggestion to J\"org Nonnenmacher).  Since that
time, a number of measurement tools have been released to the
public in the form of screen savers or daemons.
\dfn{Grenouille}~\cite{grenouille}, which is used for measuring
available bandwidth in DSL connections, was perhaps the first, and
appears to be the most widely adopted.  More recently, we have
seen the introduction of \dfn{NETI@home}~\cite{neti}, a passive
measurement tool inspired by the distributed signal analysis tool,
\dfn{SETI@home}~\cite{seti}.  In the summer of 2004, the first
tracerouting tool was made available: \dfn{DIMES}~\cite{dimes}
conducts traceroutes and pings from, at the time of this writing,
323 sites in 43 countries.

Given that much large scale network mapping is on the way,
contemplating such a measurement system demands attention to
efficiency, in order to avoid generating undesirable network load.
Unfortunately, this issue has not been yet successfully tackled by
the research community.  As Cheswick, Burch and Branigan note,
such a system ``would have to be engineered very carefully to
avoid abuse''~\cite[Sec.~7]{mapping}. Traceroutes emanating from a
large number of monitors and converging on selected targets can
easily appear to be a DDoS attack. Whether or not it triggers
alarms, it clearly is not desirable for a measurement system to
consume undue network resources.  A \dfn{traceroute@home} system,
as we label this class of applications, must work hard to avoid
sampling router interfaces and traversing links multiple times,
and to avoid multiple pings of end systems.

This lack of consideration on efficiency is in contrast to the
number of papers on efficient monitoring of networks that are in a
single administrative domain (see for instance, Bejerano and
Rastogi's work \cite{robust}). However, both problems are
completely different.  An administrator knows their entire network
topology in advance, and can freely choose where to place their
monitors. Neither of these assumptions hold for monitoring the
internet with a highly distributed software.  Since the existing
literature is based upon these assumptions, we need to look
elsewhere for solutions.

In this paper, we first evaluate the extent to which classical
topology discovery systems involve duplicated effort. By classical
topology discovery, we mean those tracerouting from a small number
of monitors to a large set of common destinations, such as
skitter.  Duplicated effort in such systems takes two forms:
measurements made by an individual monitor that replicate its own
work, and measurements made by multiple monitors that replicate
each other's work. We term the first \dfn{intra-monitor
redundancy} and the second \dfn{inter-monitor redundancy}.

Using skitter data from August 2004, we quantify both kinds of
redundancy.  We show that intra-monitor redundancy is high close
to each monitor.  This fact is not surprising given the tree-like
structure (or \dfn{cone} \cite{connectivity}) of routes emanating
from a single monitor.  However, the degree of such redundancy is
quite serious: some interfaces are visited once for each
destination probed (which could be hundreds of thousands of times
per day in a large-scale system).  Further, with respect to
inter-monitor redundancy, we find that most interfaces are visited
by all monitors, especially when close to destinations.  This
latter form of redundancy is also potentially quite serious, since
this would be expected to grow proportional to the number of
monitors in future large-scale measurement systems.

Our analysis of the nature of redundant probing suggests more
efficient algorithms for topology discovery.  In particular, our
second contribution is to propose and evaluate an algorithm called
Doubletree.  We show that Doubletree can dramatically reduce the
impact on routers and final destinations by reducing redundant
probing, while maintaining high coverage in terms of interface and
link discovery.  Doubletree is particularly effective at removing
the worst cases of highly redundant probing that would be expected
to raise alarms.

Doubletree takes advantage of the tree-like structure of
single-source or single-destination routing to avoid duplication
of effort. Unfortunately, general strategies for reducing these
two kinds of redundancy are in conflict. On the one hand,
intra-monitor redundancy is reduced by starting probing far from
the monitor, and working backward along the tree-like structure
that is rooted at that monitor.  Once an interface is encountered
that has already been discovered by the monitor, probing stops. On
the other hand, inter-monitor redundancy reduced by probing
forwards towards a destination until encountering a
previously-seen interface.  In this case, the tree-like structure
is based on the probes of multiple monitors towards a same
destination.

We show how to balance these conflicting strategies in Doubletree.
In Doubletree, probing starts at a distance that is intermediate
between monitor and destination.  We demonstrate methods for
choosing this distance, and we then evaluate the resulting
performance of Doubletree. Despite the challenge inherent in
reducing both forms of redundancy simultaneously, we show that
probing via Doubletree can reduce measurement load by
approximately 70\% while maintaining interface and link coverage
above 90\%.

The Doubletree algorithm requires communication between monitors
in order to reduce inter-monitor redundancy.  Information
regarding interfaces seen when tracing towards each destination
must be shared.  However, this can lead to considerable overhead
as the number of known interfaces grows. In this paper, we also
propose to reduce this cost through the use of Bloom filters for
lossy encoding of the interface set. Surprisingly, we find that
using Bloom filters can increase node and link coverage without a
large increase in redundancy.

The remainder of this paper is organized as follow:
Chapter~\ref{redundancy} evaluates the extent of redundancy in
classical topology tracing systems. Chapter~\ref{algo} describes
and evaluates the Doubletree algorithm.  Chapter~\ref{bf} shows
how Bloom filters can help to reduce the communication cost
required by our algorithm. Finally, Chapter~\ref{conclusion}
concludes this paper and discusses directions for future work.

\section{Redundancy}\label{redundancy}

In this chapter we quantify and analyze the extensive measurement
redundancy that can be found in a classical topology discovery
system.

\subsection{Methodology}\label{redundancy.methodology}

Our study is based on skitter data from August \nth{1} through
\nth{3}, 2004. This data set was generated by 24 monitors located
in the United States, Canada, the United Kingdom, France, Sweden,
the Netherlands, Japan, and New Zealand. The monitors share a
common destination set of nearly one million IPv4 addresses. Each
monitor cycles through the destination set at its own rate, taking
typically three days to complete a cycle. For the purpose of our
studies, in order to reduce computing time to a manageable level,
we worked from a limited destination set of 50,000, randomly
chosen from the original set.

Visits to host and router interfaces are the metric by which we
evaluate redundancy.  We consider an interface to have been
visited if its IP address appears at one of the hops in a
traceroute.  Though it would be of interest to calculate the load
at the host and router level, rather than at the individual
interface level, we make no attempt to disambiguate interfaces in
order to obtain a router-level graph.  The alias resolution
techniques described by Pansiot and Grad~\cite{onRoutes}, by
Govindan and Tangmunarunkit~\cite{heuristics}, for
\emph{Mercator}, and applied in the \emph{iffinder} tool from
\textsc{Caida}~\cite{iffinder}, would require active probing
beyond the skitter data, preferably at the same time that the
skitter data is collected. The methods used by Spring et
al.~\cite{rocketfuel}, in \emph{Rocketfuel}, and by Teixeira et
al.~\cite{pathDiversity}, apply to routers in the network core,
and are untested in stub networks.  Despite these limitations, we
believe that the load on individual interfaces is a useful
measure. As Broido and claffy note~\cite{connectivity},
``interfaces are individual devices, with their own individual
processors, memory, buses, and failure modes. It is reasonable to
view them as nodes with their own connections.''

What does it mean for an IP address to appear at a given hop
distance from a monitor?  Skitter, like many standard traceroute
implementations, sends three probe packets for each hop count. Our
accounting assumes a baseline probing method which, instead, tries
up to three times to get a response at each hop. After the first
successful response, the probe moves to the next hop.  Thus, the
first successfully reached address at each hop is the one used. If
none of the three probes are returned, the hop is recorded as
non-responding.  In terms of redundancy, this method in fact
revisits interfaces less often than the current version of
skitter, but is more consistent with the goal of minimizing
measurement load, and its behavior can be easily simulated from
skitter traces.

Even if an IP address is returned for a given hop count, it might
not be valid.  Due to the presence of poorly configured routers
along traceroute paths, skitter occasionally records anomalies
such as private IP addresses that are not globally routable.  We
account for invalid hops as if they were non-responding hops.  The
addresses that we consider as invalid are a subset of the
special-use IPv4 addresses described in RFC~3330~\cite{rfc3330}.
Specifically, we eliminate visits to the private IP address blocks
10.0.0.0/8, 172.16.0.0/12, and 192.168.0.0/16.  We also remove the
loopback address block 127.0.0.0/8.  In our data set, we find
4,435 different special addresses, more precisely 4,434 are
private addresses and only one is a loopback address.  Special
addresses cover around 3\% of the entire considered addresses set.
Though there were no visits in the data to the following address
blocks, they too would be considered invalid: the ``this network''
block 0.0.0.0/8, the 6to4 relay anycast address block
192.88.99.0/24, the benchmark testing block 198.18.0.0/15, the
multicast address block 224.0.0.0/4, and the reserved address
block formerly known as the Class E addresses, 240.0.0.0/4, which
includes the \textsc{lan} broadcast address, 255.255.255.255.

We evaluate the redundancy at two levels. One is the microscopic
level of a single monitor, considered in isolation from the rest
of the system.  This intra-monitor redundancy is measured by the
number of times the same monitor visits an interface.  The other,
macroscopic, level considers the system as an ensemble of
monitors. This inter-monitor redundancy is measured by the number
of monitors that visit a given interface, counting only once each
monitor that has non-zero intra-monitor redundancy for that
interface. By separating the two levels, we separate the problem
of redundancy into two problems that can be treated somewhat
separately.  Each monitor can act on its own to reduce its
intra-monitor redundancy,  but cooperation between monitors is
required to reduce inter-monitor redundancy.

\subsection{Description of the Plots}\label{redundancy.description}

Since the redundancy distributions are generally skewed, quantile
plots give us a better sense of the data than would plots of the
mean and variance. There are several possible ways to calculate
quantiles.  We calculate them in the manner described by
Jain~\cite[p.~194]{jainArtCSPerfAnalysis}, which is: rounding to
the nearest integer value to obtain the index of the element in
question, and using the lower integer if the quantile falls
exactly halfway between two integers.

\begin{figure}[tbp]
  \begin{center}
    \includegraphics[height=2.5cm]{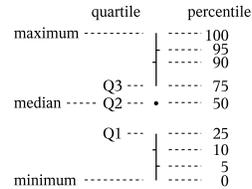}
  \end{center}
  \caption{Quantiles key}
  \label{quantiles.key}
\end{figure}

Fig.~\ref{quantiles.key} provides a key to reading the quantile
plots found in Figs.~\ref{redundancy.intra.fig} and
\ref{redundancy.inter.global} and figures found later in the
paper.

A dot marks the median (the \nth{2} quartile, or \nth{50}
percentile). The vertical line below the dot delineates the range
from the minimum to the \nth{1} quartile, and leaves a space from
the \nth{1} to the \nth{2} quartile. The space above the dot runs
from the \nth{2} to the \nth{3} quartile, and the line above that
extends from the \nth{3} quartile to the maximum.  Small tick
marks to either side of the lines mark some additional
percentiles: marks to the left for the \nth{10} and \nth{90}, and
marks to the right for the \nth{5} and \nth{95}.

In the case of highly skewed distributions, or distributions drawn
from small amounts of data, the vertical lines or the spaces
between them might not appear. For instance, if there are tick
marks but no vertical line above the dot, this means that the
\nth{3} quartile is identical to the maximum value. In the
figures, each quantile plot sits directly above an accompanying
bar chart that indicates the quantity of data upon which the
quantiles were based. For each hop count, the bar chart displays
the number of interfaces at that distance.  For these bar charts,
a log scale is used on the vertical axis. This allows us to
identify quantiles that are based upon very few interfaces (fewer
than twenty, for instance), and so for which the values risk being
somewhat arbitrary.

\subsection{Intra-monitor Redundancy}\label{redundancy.intra}

Intra-monitor redundancy occurs in the context of the tree-like
graph that is generated when all traceroutes originate at a single
point. Since there are fewer interfaces closer to the monitor,
those interfaces will tend to be visited more frequently. In the
extreme case, if there is a single gateway router between the
monitor and the rest of the internet, a single IP address
belonging to that router should show up in every one of the
traceroutes.

We measure intra-monitor redundancy by considering all traceroutes
from the monitor to the common destinations, whether there be
problems with a traceroute, as described in
Sec.~\ref{redundancy.methodology}, or not.

Having calculated the intra-monitor redundancy for each interface,
we organize the results by the distance of the interfaces from the
monitor.  We measure distance by hop count.  Since the same
interface can appear at a number of different hop counts from a
monitor, for instance if routes change between traceroutes, we
arbitrarily attribute to each interface the hop count at which it
was first visited.  This process yields, for each hop count, a set
of interfaces that we sort by number of visits.  We then plot, hop
by hop, the redundancy distribution for interfaces at that hop
count.

  \subsubsection{Results}\label{redundancy.intra.results}

  \begin{figure}[!t]
    \begin{center}
      \subfigure[\texttt{arin}]{\label{redundancy.intra.arin}
        \includegraphics[width=5.5cm]{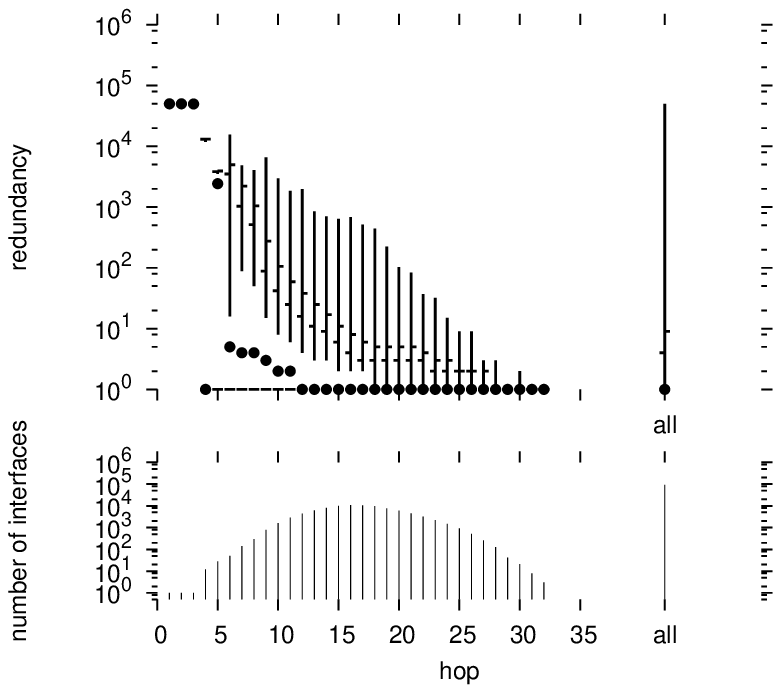}}
      \subfigure[\texttt{champagne}]{\label{redundancy.intra.champ}
        \includegraphics[width=5.5cm]{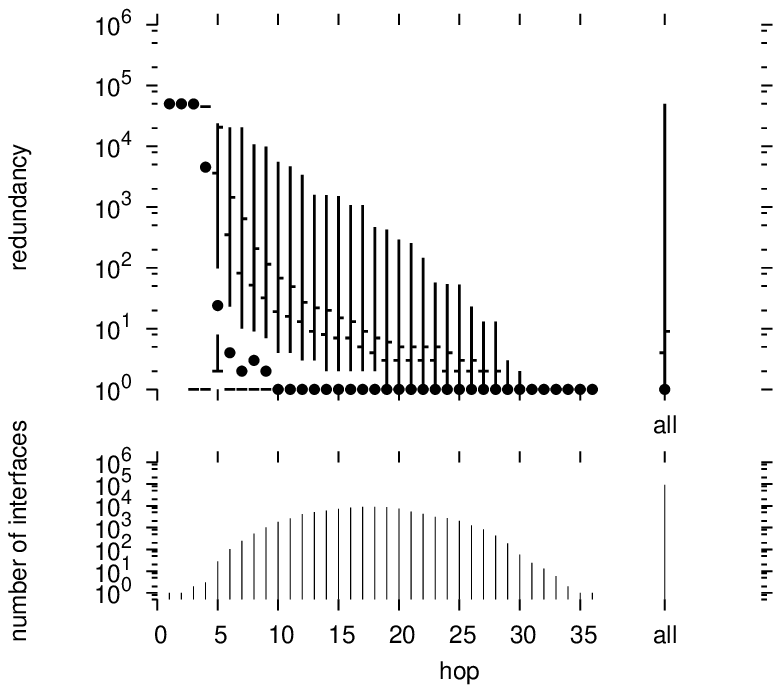}}
    \end{center}
    \caption{Skitter intra-monitor redundancy}
    \label{redundancy.intra.fig}
  \end{figure}

Fig.~\ref{redundancy.intra.fig} shows intra-monitor redundancy quantile
distributions for two representative skitter monitors: \url{arin} and
\url{champagne}.

Looking first at the histograms for interface counts (lower half
of each plot), we see that these data are consistent with
distributions typically seen in such cases.  Plotted on a linear
scale (not shown here) these distributions display the familiar
bell-shaped curve typical of internet interface distance
distributions.  The distribution for \url{champagne} is fairly
typical of all monitors. It represents the 92,355 unique IP
addresses discovered by that monitor.  This value is shown as a
separate bar to the right of the histogram, labeled ``all''.  We
see that the interface distances are distributed with a mean at 18
hops corresponding to a peak of 9,135 interfaces that are visited
at that distance.

The quantile plots show the nature of the intra-monitor redundancy
problem. Looking first to the bar at the right hand of each chart,
showing the quantiles for all of the interfaces taken together, we
can see that the distributions are highly skewed. The median
interface has a redundancy of one.  Even the \nth{75} quantile is
one, as evidenced by the lack of a gap between the dot and the
line representing the top quarter of values. However, for a very
small portion of the interfaces there is a very high redundancy.
The maximum redundancy in each case is 50,000---equal to the
number of destinations.

Looking at how the redundancy varies by distance, we see that the
problem is worse the closer one is to the monitor.  This is what
we expect given the tree-like structure of routing from a monitor,
but here we see how serious the phenomenon is from a quantitative
standpoint. For the first three hops from each monitor, the median
redundancy is 50,000. A look at the histograms shows that there
are very few interfaces at these distances.  Just one interface
for \url{arin}, and the same for \url{champagne}, save for the
presence of a second interface at the third hop. This second
interface is only visited once, as represented by the presence of
the \nth{5} and \nth{10} percentile marks (since there are only
two data points, the lower valued point is represented by the
entire lower quarter of values on the plot).

Beyond three hops, the median redundancy drops rapidly.  By the
sixth hop, in both cases, the median is below ten.  By the twelfth
hop, the median is one.  However, the distributions remain highly
skewed.  Even fifteen hops out, some interfaces experience a
redundancy on the order of several hundred visits.  With small
variations, these patterns are repeated for each of the monitors.

From the point of view of planning a measurement system, the
extreme values are the most worrisome.  It is clear that there is
significant duplicated effort, but it is especially concentrated
in selected areas.  The problem is most severe on the first few
interfaces, but even interfaces many hops out receive hundreds or
thousands of repeat visits.  Beyond the danger of triggering
alarms, there is a simple question of measurement efficiency.
Resources devoted to reprobing the same interfaces would be better
saved, or reallocated to more fruitful probing tasks.

  \begin{table}[!t]
    \begin{center}
      \begin{tabular}{lr@{.}l}
        \multicolumn{2}{c}{Destinations}\\
        Responding     & 59 & 7\%\\
        Not responding & 40 & 3\%\\
        \multicolumn{2}{c}{Probes}\\
        Interface discovery   & 10 & 4\%\\
        Invalid addresses     &  1 & 5\%\\
        No response           &  0 & 5\%\\
        Redundant             & 87 & 6\%\\
      \end{tabular}
    \end{center}
    \caption{Probes statistics for \texttt{champagne}}
    \label{redundancy.intra.tab}
  \end{table}

Table~\ref{redundancy.intra.tab} presents additional statistics for
\url{champagne}.  The first part of the table indicates the portion of
destinations that respond and the portion that do not respond. Fully 40.3\%
of the traceroutes do not terminate with a destination response.  The second
part of the table describes redundancy in terms of probes sent, rather than
from an interface's perspective. Only 10.4\% of probes serve to discover a
new interface.  (Note: in the intra-monitor context, an interface is
considered to be new if that particular monitor has not previously visited
it.)  An additional 2.0\% of probes hit invalid addresses, as defined in
Sec.~\ref{redundancy.methodology}, or do not result in a response.  This
leaves 87.6\% of the probes that are redundant in the sense that they visit
interfaces that the monitor has already discovered.  The statistics in this
table are typical of the statistics for every one of the 24 monitors.

\subsection{Inter-monitor Redundancy}\label{redundancy.inter}

Inter-monitor redundancy occurs when multiple monitors visit the
same interface. The degree of such redundancy is of keen interest
to us when increasing the number of monitors by several orders of
magnitude is envisaged.

We calculate the inter-monitor redundancy for each interface by
counting the number of monitors that have visited it.  A monitor
can be counted at most once towards an interface's inter-monitor
redundancy, even if it has visited that interface multiple times.
For a given interface, the redundancy is calculated just once with
respect to the entirety of the monitors: it does not vary from
monitor to monitor as does intra-monitor redundancy. However, what
does vary depending upon the monitor is whether the particular
interface is seen, and at what distance.  In order to attribute a
single distance to an interface, a distance that does not depend
upon the perspective of a single monitor but that nonetheless has
meaning when examining the effects of distance on redundancy, we
attribute the minimum distance at which an interface has been seen
among all the monitors.

  \subsubsection{Results}\label{redundancy.inter.results}
  \begin{figure}[tbp]
    \begin{center}
      \includegraphics[width=5cm]{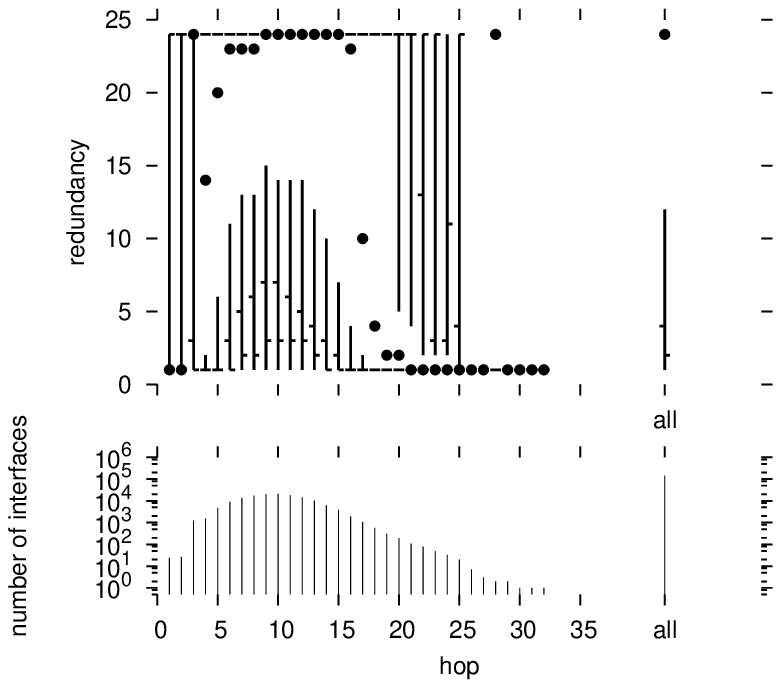}
    \end{center}
    \caption{Skitter inter-monitor redundancy}
    \label{redundancy.inter.global}
  \end{figure}

Fig.~\ref{redundancy.inter.global} shows inter-monitor redundancy
for the skitter data.

The distribution of interfaces by hop count differs from the
intra-monitor case due to the difference in how we account for
distances.  The mean is closer to the traceroute source (10 hops),
corresponding to the peak of 21,222 interfaces that are visited at
that distance.

The redundancy distribution also has a very different aspect.
Considering, first of all, the redundancy over all of the
interfaces (at the far right of the plot), we see that the median
interface is visited by all 24 monitors, which is a subject of
great concern. The distribution is also skewed, though the effect
is less dramatic since the vertical axis is a linear scale, with
only 24 possible values.

We also see a very different distribution by distance.  Interfaces
that are very close in to a monitor, at one or two hops, have a
median inter-monitor redundancy of one.  The same is true of
interfaces that are far from all monitors, at distances over 20,
though there are very few of these.  (The presence of an interface
at hop 27 that is seen by all monitors serves to raise the median
at that distance to 24.)  What is especially notable is that
interfaces at intermediate distances (6 to 15) tend to be visited
by all, or almost all, of the monitors.  Though their distances
are in the middle of the distribution, this does not mean that the
interfaces themselves are in the middle of the network.  Many of
these interfaces are in fact destinations.  Recall that every
destination is targeted by every host.

\section{Algorithm}\label{algo}

In this chapter, we present the Doubletree algorithm, our method
for probing the network in a friendly manner while discovering
nearly all the interfaces and links that a classical tracerouting
approach would discover.

Sec.~\ref{algo.description} describes how Doubletree works.
Sec.~\ref{algo.tension} discusses the results of varying the
single parameter of this algorithm. Finally,
Sec.~\ref{algo.redundancy} shows the extent of intra- and
inter-monitor redundancy reduction when using the algorithm.

\subsection{Description}\label{algo.description}

\begin{figure}[!tbp]
    \begin{center}
      \subfigure[Monitor-rooted]{\label{treeStructureFig.intra}
        \includegraphics[width=5cm]{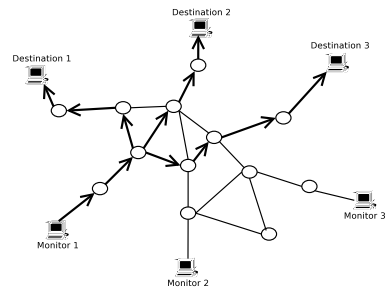}}
      \subfigure[Destination-rooted]{\label{treeStructureFig.inter}
        \includegraphics[width=5cm]{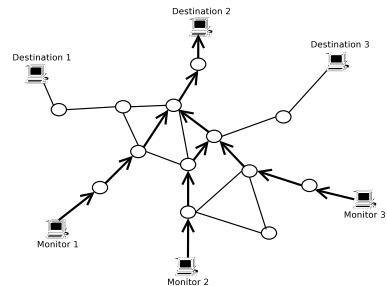}}
    \end{center}
    \caption{Tree-like routing structures}
    \label{treeStructureFig}
  \end{figure}

Doubletree takes advantage of the tree-like structure of routes in
the internet. Routes lead out from a monitor towards multiple
destinations in a tree-like way, as shown in
Fig.~\ref{treeStructureFig.intra}, and the manner in which routes
converge towards a destination from multiple monitors is similarly
tree-like, as shown in Fig.~\ref{treeStructureFig.inter}.  The
tree is an idealisation of the structure encountered in practice.
Paths separate and reconverge.  Loops can arise.  But a tree may
be a good enough first approximation on which to base a probing
algorithm.

A probing algorithm can reduce its redundancy by tracking its
progress through a tree, as it probes from the direction of the
leaves towards the root.  So long as it is probing in a previously
unknown part of the tree, it continues to probe.  But once it
encounters a node that is already known to belong to the tree, it
stops.  The idea being that the remainder of the path to the root
must already be known.  In reality, there is only a likelihood and
not a certainty that the remainder of the path is known.  The
redundancy saved by not reprobing the same paths to the root may
nonetheless be worth the loss in coverage that results from not
probing the occasional different path.

Doubletree uses both the monitor-rooted and the destination-rooted
trees. When probing backwards from the destinations towards the
monitor, it applies a stopping rule based upon the monitor-rooted
tree. The goal in this case is to reduce intra-monitor redundancy.
When probing forwards, the stopping rule is based upon the
destination-rooted tree, with the goal being to reduce
inter-monitor redundancy. There is an inherent tension between the
two goals.

Suppose the algorithm were to start probing only far from each
monitor. Probing would necessarily be backwards.  In this case,
the destination-based trees cannot be used to reduce redundancy. A
monitor might discover, with destination $d$ at hop $h$, an
interface that another monitor also discovered when probing with
destination $d$.  However this does not inform the monitor as to
whether the interface at hop $h-1$ is likely to have been
discovered as well.  So it is not clear how to reduce
inter-monitor redundancy when conducting backwards probing.

Similarly, when conducting forwards probing (of the classic
traceroute sort), it is not clear how intra-monitor redundancy can
be avoided.  Paths close to the monitor will tend to be probed and
reprobed, for lack of knowledge of where the path to a given
destination might diverge from the paths already seen.

In order to reduce both inter- and intra-monitor redundancy,
Doubletree starts probing at what is hoped to be an intermediate
point.  For each monitor, there is an initial hop count $h$.
Probing proceeds forwards from $h$, to $h+1$, $h+2$, and so forth,
applying the stopping rule based on the destination-rooted tree.
Then it probes backwards from $h$, to $h-1$, $h-2$, etc., using
the monitor-based tree stopping rule.  In the special case where
there is no response at distance $h$, the distance is halved, and
halved again until there is a reply, and probing continues
forwards and backwards from that point.

Rather than maintaining detailed information on tree structures,
it is sufficient for the stopping rules to make use of sets of
interfaces.  Each monitor tracks the interfaces that it has
discovered.  These form a stop set $B$, called the \dfn{backwards
tracing stop set}, or more concisely, the \dfn{local stop set}, to
be used in a monitor's own backwards probing. When probing
backwards from a destination $d$, encountering an interface in $B$
causes the monitor to stop and move on to the next destination.
Each monitor also receives another stop set, $F$, called the
\dfn{forwards tracing stop set}, or more concisely, the
\dfn{global stop set}, that contains
$(\mathrm{interface},\mathrm{destination})$ pairs. When probing
forwards towards destination $d$ and encountering an interface
$i$, forwards probing stops if $(i,d) \in F$. Communication
between monitors is needed in order to share this second stop set.

Only one aspect of Doubletree has been suggested in prior
literature. Govindan and Tangmunarunkit~\cite{heuristics} employ
backwards probing with a stopping rule in the Mercator system, in
order to reduce intra-monitor redundancy.  However, no results
have been published regarding the efficacy of this approach.  Nor
have the effects on inter-monitor redundancy been considered, or
the tension between reducing the two types of redundancy (for
instance, Mercator tries to start probing at the destination, or
as close to it as possible). Nor has any prior work suggested a
manner in which to exploit the tree-like structure of routes that
converge on a destination.  Finally, no prior work has suggested
cooperation among monitors.

Algorithm~\ref{algo.formal} is a formal definition of the
Doubletree algorithm. It assumes that the following two functions
are defined. The \textit{response}() procedure returns true if an
interface replies to at least one of the probes that were sent.
\textit{halt}() is a primitive that checks if the probing must be
stopped for different reasons: a loop is detected or a gap (five
successive non-responding nodes) is discovered.

  \begin{algorithm}[!t]
    \caption{Doubletree}
    \label{algo.formal}
    \begin{algorithmic}[1]
      \Require $F$, the global stop set received by this monitor.
      \Ensure $F$ updated with all (interface,destination) pairs discovered by this monitor.
      \Statex
      \Procedure{Doubletree}{$h$, $D$}
          \State $B \leftarrow \emptyset$\Comment{Local stop set}
          \ForAll{$d \in D$}\Comment{Destinations}
              \State $h \leftarrow$
              \textsc{AdaptHValue}($h$)\Comment{Initial hop}
              \State \textsc{TraceForwards}($h$, $d$)
              \State \textsc{TraceBackwards}($h-1$, $d$)
          \EndFor
      \EndProcedure
      \Statex
      \Procedure{AdaptHValue}{$h$}
          \While{$\neg \mathrm{response}(v_h) \wedge h \neq
          1$}\Comment{$v_h$ the interface discovered at $h$ hops}
              \State $h \leftarrow \frac{h}{2}$\Comment{$h$ an integer}
          \EndWhile
          \State \textbf{return} $h$
      \EndProcedure
      \Statex
      \Procedure{TraceForwards}{$i$, $d$}
          \While{$v_i \neq d \wedge (v_i, d) \notin F \wedge \neg\mathrm{halt}()$}
              \State $F \leftarrow F \bigcup (v_i, d)$
              \State $i++$
          \EndWhile
      \EndProcedure
      \Statex
      \Procedure{TraceBackwards}{$i$, $d$}
          \While{$i \geqslant 1 \wedge v_i \notin B \wedge \neg\mathrm{halt}()$}
              \State $B = B \bigcup v_i$
              \State $F = F \bigcup (v_i, d)$
              \State $i--$
          \EndWhile
      \EndProcedure
    \end{algorithmic}
  \end{algorithm}

This algorithm has only one tunable parameter: the initial hop
count $h$.  In the remainder of this section, we explain how to
set this parameter in terms of another parameter that we call $p$.

We wish for each monitor to be able to determine a reasonable
value for $h$: one that is far enough from the monitor to avoid
excess intra-monitor redundancy, yet not so far as to generate too
much inter-monitor redundancy. Since each monitor will be
positioned differently with respect to the internet, what is
reasonable for one monitor might not be reasonable for another. We
thus base our rule for choosing $h$ on the distribution of path
lengths as seen from the perspective of a given monitor.  The
general idea is to start probing at a distance that is rich in
interfaces, but that is not so far as to exacerbate inter-monitor
redundancy.

Based upon our intra-monitor redundancy studies, discussed above,
we would expect an initial hop distance of five or more from the
typical monitor to be fairly rich in interfaces.  However, we also
know that this is the distance at which inter-monitor redundancy
can become a problem.  We are especially concerned about
inter-monitor redundancy at destinations, because this is what is
most likely to look like a DDoS attack.

  \begin{figure}[!tbp]
    \begin{center}
      \includegraphics[width=5cm]{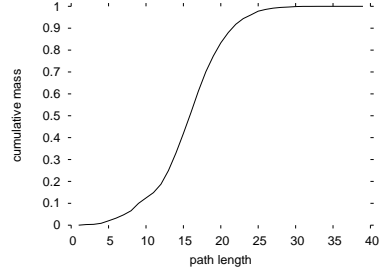}
    \end{center}
    \caption{Lengths of paths from monitor \texttt{apan-jp}}
    \label{algo.cdf}
  \end{figure}

One parameter that a monitor can estimate without much effort is
its probability of hitting a responding destination at any
particular hop count $h$.  For instance, Fig.~\ref{algo.cdf} shows
the cumulative mass plot of path lengths from monitor
\url{apan-jp}. If \url{apan-jp} choses $h=10$, that implies a
$0.1$ probability of hitting a responding destination on the first
probe.  The shape of this curve is very similar for each of the 24
skitter monitors, but the horizontal position of the curve can
vary by a number of hops from monitor to monitor.  So if we are to
fix the probability, $p$, of hitting a responding destination on
the first probe, there will be different values $h$ for each
monitor, but that value will correspond to a similar level of
incursion into the network across the board.

We have chosen $p$ to be the single independent parameter that
must be tuned to guide Doubletree. In the following section, we
study the effect of varying $p$ on the tension between inter- and
intra-monitor redundancy, and the overall interface and link
coverage that Doubletree obtains.

\subsection{Tuning the Parameter p}\label{algo.tension}

This section discusses the effect of varying $p$.
Sec.~\ref{algo.tension.methodology} describes our experimental
methodology, and Sec.~\ref{algo.tension.results} presents the
results.

\subsubsection{Methodology}\label{algo.tension.methodology}

In order to test the effects of the parameter $p$ on both
redundancy and coverage, we implement Doubletree in a simulator.
We examine the following values for $p$: between 0 (i.e., forwards
probing only) and 0.2, we increment $p$ in steps of 0.01. From 0.2
to $1$ (i.e., backwards probing in all cases when the destination
replies to the first probe), we increment $p$ in steps of 0.1. As
will be shown, the concentration of values close to 0 allows us to
trace the greater variation of behavior in this area.

To validate our results, we run the simulator using the same
skitter data set we considered in Sec.~\ref{redundancy}.  We
assume that Doubletree is running on the skitter monitors, during
the same period of time that the skitter data represents, and
implementing the same baseline probing technique described in
Sec.~\ref{redundancy.methodology}, of probing up to three times at
a given hop distance.  The difference lies in the order in which
Doubletree probes the hops, and the application of Doubletree's
stopping rules.

Doubletree requires communication of the global stop set from one
monitor to another.  We therefore choose a random order for the
monitors and simulate the running of Doubletree on each one in
turn.  The global stop set is added to and passed on to each
monitor in turn.  This is a simplified scenario compared to the
way in which a fully operational cooperative topology discovery
protocol might function, which is to say with all of the monitors
probing and communicating in parallel. However, we feel that the
scenario allows greater realism in the study of intra-monitor
redundancy. The typical monitor in a large, highly distributed
infrastructure will begin its probing in a situation in which much
of the topology has already been discovered by other monitors. The
closest we can get to simulating the experience of such a monitor
is by studying what happens to the last in our random sequence of
monitors.  All Doubletree intra-monitor redundancy results are for
the last monitor in the sequence.  (Inter-monitor redundancy, on
the other hand, is monitor independent.)

\subsubsection{Results}\label{algo.tension.results}

  \begin{figure}[!tbp]
    \begin{center}
      \includegraphics[width=8cm]{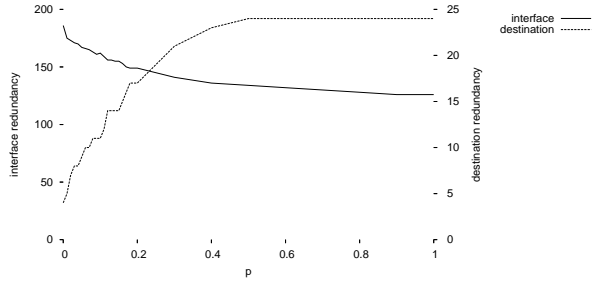}
    \end{center}
    \caption{Doubletree redundancy, \nth{95} percentile.  Inter-monitor redundancy on destinations,
      gross redundancy on router interfaces.}
    \label{algo.tension.redundancy}
  \end{figure}

Since the value $p$ has a direct effect on the redundancy of
destination interfaces, we initially look at the effect of $p$
separately on destination redundancy and on router interface
redundancy.  We are most concerned about destination redundancy
because of its tendency to appear like a DDoS attack, and we are
concerned in particular with the inter-monitor redundancy on these
destinations, because a variety of sources is a prime indicator of
such an attack.  The right-side vertical axis of
Fig.~\ref{algo.tension.redundancy} displays destination
redundancy.  With regards router interface redundancy, we are
concerned with overall load, and so we consider a combined intra-
and inter-monitor redundancy measure that we call \dfn{gross
redundancy}, that counts the total number of visits to an
interface.  For both destinations and router interfaces, we are
concerned with the extreme values, so we consider the \nth{95}
percentile.

As expected the \nth{95} percentile inter-monitor redundancy on
destinations increases with $p$. Values increase until $p=0.5$, at
which point they plateau at 24.  The point $p=0.5$ is the point at
which, in 50\% of the cases, the probe sent to a distance $h$ hits
a destination.  Doubletree allows a reduction in \nth{95}
percentile inter-monitor redundancy when compared to classical
probing for lower values of $p$.  The reduction is 84\% when
$p=0$.

As opposed to destination redundancy, the \nth{95} percentile
gross router interface redundancy decreases with $p$.  The
\nth{95} percentile for the internal interface gross redundancy
using the classical approach is 449. Doubletree thus allows a
reduction between 59\% ($p=0$) and 72\% ($p=1$).

This preliminary analysis suggests that Doubletree should employ a
low value for $p$, certainly below 0.5, in order to reduce
inter-monitor redundancy on destinations.  This is a very
different approach than that taken by Mercator, which attempts to
hit a destination every time.  On the other hand, too low a value
will have a negative impact on router interfaces.  We now examine
other evidence that will bear on our choice of $p$.

  \begin{figure}[!tbp]
    \begin{center}
      \includegraphics[width=8cm]{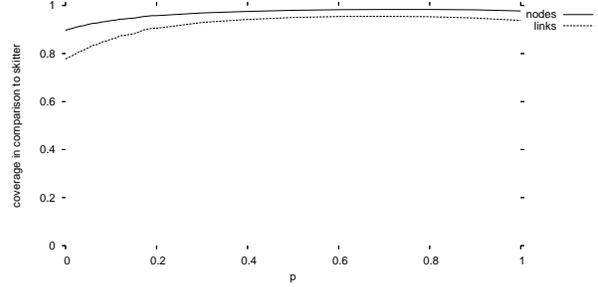}
    \end{center}
    \caption{Links and nodes coverage in comparison to classic probing}
    \label{algo.tension.coverage}
  \end{figure}

Fig.~\ref{algo.tension.coverage} illustrates the effects of $p$ on
the node and link coverage percentage in comparison to classic
probing. As we can see, the coverage increases with $p$ but a
small decrease is noticed for values of $p$ greater than $0.7$.
The maximum of coverage is reached when $p=0.7$: Doubletree
discovers 95,49\% of links and 98,4\% of nodes.  The minimum of
coverage appears when $p=0$: 77\% of links and 89\% of nodes.
However, link coverage grows rapidly for $p$ values between $0$
and $0.4$. After that point, a kind of plateau is reached, before
a small decrease.

Fig.~\ref{algo.tension.coverage} shows that the information (i.e.
links and nodes) discovery of our algorithm is satisfactory,
especially for non zero values of $p$.

  \begin{figure}[!tbp]
    \begin{center}
      \includegraphics[width=8cm]{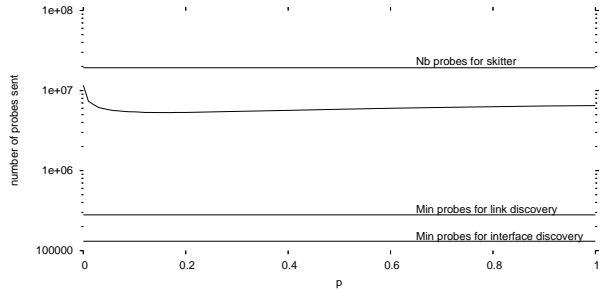}
    \end{center}
    \caption{Amount of probes sent}
    \label{algo.tension.visit}
  \end{figure}

Fig.~\ref{algo.tension.visit} shows the effects of $p$ on the
number of probes sent.  The horizontal axis indicates the value
for $p$.  The vertical axis represents the number of probes sent.
If we consider an ideal system in which each probe sent visits a
new interface (i.e., there is no redundancy), the number of probes
sent to discover all the interfaces will be 131,078 (i.e., the
number of different interfaces in the data set).  The lowest
vertical bar represents this fact. Furthermore, if the ideal
system is also able to elicit links without any redundancy, then
it should send 279,798 probes. The second vertical line considers
that point in the plot.  On the other hand, if our system works
like skitter, it has to send 19,280,551 probes.  This is
represented by the highest vertical line.  In order to plot all
these lines on the same figure, the vertical axis has been plotted
in a log scale.  With Doubletree, the number of probes needed
varies between 11,684,439 (i.e., a reduction of 40\% in comparison
to the classical approach) and 5,330,098 (i.e., a reduction of
73\% in comparison to the classical approach).   This minimum is
reached when $p=0.12$.

In the skitter data set that we consider, not all nodes (internal
interfaces and destinations) necessarily reply to probes. An
internal interface might not response to probes because the router
does not send ICMP messages, or is too busy. Usually, destinations
do not reply because of security policy. In our study, we consider
two kinds of nodes that do not reply to probes: the
\dfn{non-responding nodes} and the \dfn{unidentifiable nodes}.
Non-responding nodes appear when a node does not response in the
path, but there are other interfaces (either router or
destination) that respond at a more distant hop count.  On the
other hand, unidentifiable nodes appear at the end of the path,
when skitter does not complete a path. As we do not know if these
nodes are destinations or not, we consider them to be
unidentifiable.

Table~\ref{knownUnknown} compares Doubletree with classic probing
as concerns the non-responding and unidentifiable nodes.  We can
show that we strongly reduce the impact on unidentifiable nodes.
We note that when $p$ is at its maximum, the stress on
unidentifiable nodes is identical to the classical approach.

  \begin{table}[!tbp]
    \begin{center}
      \begin{tabular}{l|cc}
        & Nonresponding & Unidentifiable\\
        \hline
        classic  & 126,168 & 512,764\\
        \hline
        $p=0$    & 34,867 & 62,009\\
        $p=0.05$ & 26,136 & 127,917\\
        $p=0.10$ & 28,857 & 162,845\\
        $p=0.15$ & 31,792 & 196,220\\
        $p=0.20$ & 34,624 & 232,215\\
        $p=0.50$ & 47,362 & 383,000\\
        $p=1$    & 52,422 & 512,764\\
      \end{tabular}
    \end{center}
    \caption{Load on anonymous interfaces}
    \label{knownUnknown}
  \end{table}

Out of concern that our solution might be too tightly tied to fit
to our data set, we perform the same experiment on another data
set of 50,000 destinations, randomly chosen from the whole set.
There is no overlapping between the two destination subsets (i.e.,
they are totally disjoint).  We find that the results obtained
with the second data set are consistent with the first one.

These results presented in this section are important in the case
of a highly distributed measurement tool.  They demonstrate that
it is possible to probe in a network friendly manner while
maintaining a very high level of topological information gathered
by monitors.

In this section, we discussed the effects of different $p$ values.
However, results permit now to identify a range of values where a
good compromise between redundancy reduction and high level of
coverage is possible.   Thus, hitting a destination with the very
first probe in 20\% of the cases seems to us to be a reasonable
maximum.  Further, in terms of coverage, a probability $p$ of
$0.05$ seems also reasonable.

In the perspective of a real system implementing our algorithm,
the value $p$ (and the corresponding $h$) cannot be chosen a
priori, as we did in our experimentations.  However, it can be
easily computed on the fly by using an iterative process, as the
monitor's knowledge about paths and topology improves.

\subsection{Redundancy Reduction}\label{algo.redundancy}

In this section, we study the effects of Doubletree on the intra-
and inter-monitor redundancy for some values of $p$.

Sec.~\ref{algo.redundancy.methodology} describes our methodology.
Sec.~\ref{algo.redundancy.intra} presents the intra- and
Sec.~\ref{algo.redundancy.inter} the inter-monitor redundancy
reduction.

\subsubsection{Methodology}\label{algo.redundancy.methodology}

We use the simulator to study the effects of Doubletree on intra-
and inter-monitor redundancy.  Again, for comparison reasons, we
use the same data set as in Sec.~\ref{redundancy}.

The plots are presented in the same way as in
Sec.~\ref{redundancy}. However, the lower part of the graphs, the
histograms, contains additional information.  The bars are now
enveloped by a curve. This curve indicates, for each hop, the
quantity of nodes discovered while using the classical method.
The bars themselves describe the number of nodes discovered by
Doubletree. Therefore, the space between the bars and the curve
represents the quantity of nodes Doubletree misses.

In Sec.~\ref{algo.tension}, we identify the range of $p$ values for which
redundancy is sufficiently low and coverage high enough. We run simulations
for $p=0.05$, $p=0.1$, $p=0.15$ and $p=0.2$ and study the effects on inter-
and intra-monitor redundancy reduction. However, we note that the
differences between the results for each $p$ value are small. Therefore, we
choose to present in the following sections only the results for $p=0.05$.

\subsubsection{Intra-monitor}\label{algo.redundancy.intra}

  \begin{figure}[!tbp]
    \begin{center}
      \includegraphics[width=5cm]{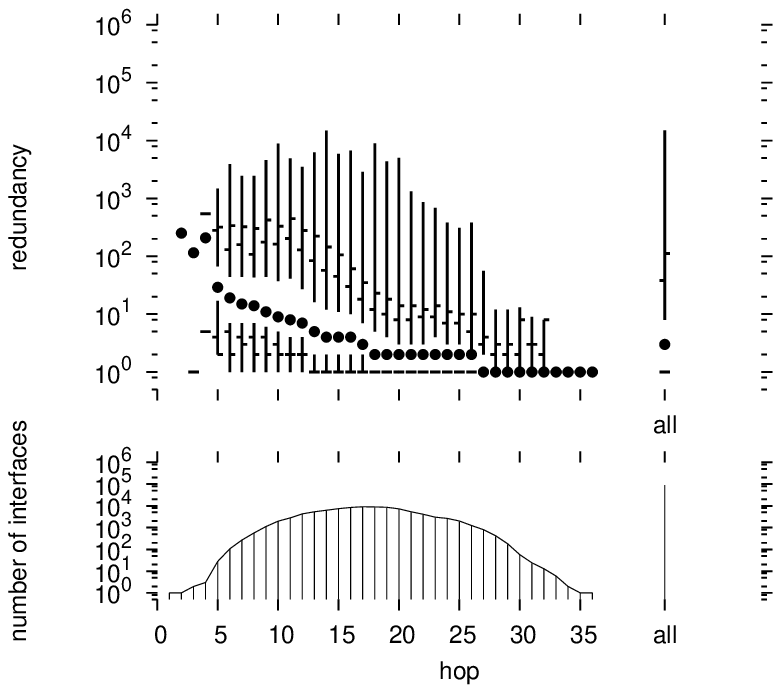}
    \end{center}
    \caption{Intra-monitor redundancy for the \texttt{champagne}
      monitor with $p=0.05$.}
    \label{algo.redundancy.intra.fig}
  \end{figure}

Fig.~\ref{algo.redundancy.intra.fig} shows intra-monitor redundancy when
using Doubletree with $p=0.05$ for a representative monitor: \url{champagne}.

First of all, we could note that, using Doubletree,
\url{champagne} is able to elicit 97\% of the interfaces in
comparison to the classical method.

Looking to the right part of the plot first, we note that the
median has a redundancy of 3.  It is a little bit higher than for
the classical method.  As in the classical approach, for a very
small number of interfaces there is a high redundancy.
Nevertheless, the maximum is 15,029. Compared to the 50,000 in the
classical approach, there is a reduction of 70\%.

Looking now at how the redundancy varies by distance, we note a
strong reduction for the median values close to the monitor.
However, for further hops, the median values drop lower than in
the classical approach (see, for comparison,
Fig.~\ref{redundancy.intra.champ}). Finally, we note that high
quantiles for hops far from the source have higher values than for
the classical method.

\subsubsection{Inter-monitor}\label{algo.redundancy.inter}

  \begin{figure}[!t]
    \begin{center}
      \includegraphics[width=5cm]{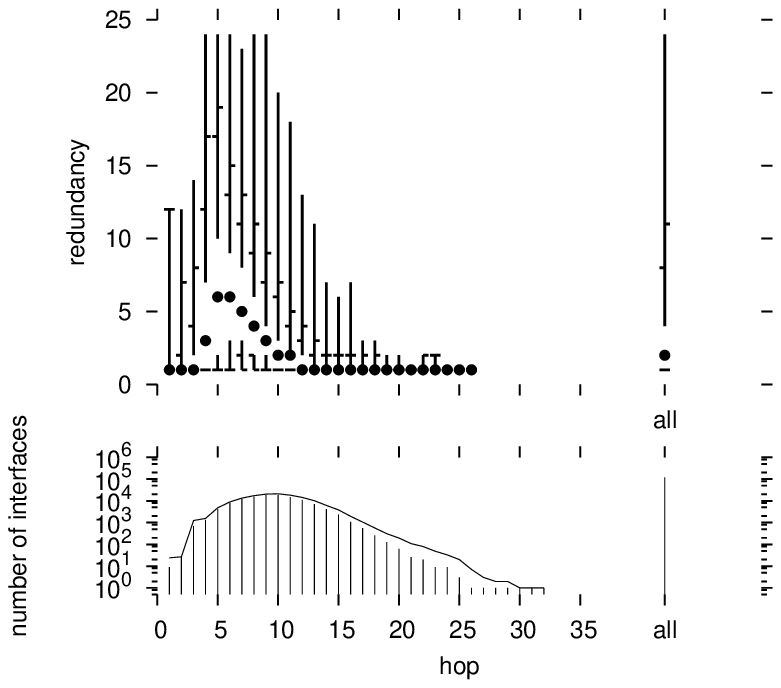}
    \end{center}
    \caption{Inter-monitor redundancy with $p=0.05$.}
    \label{algo.redundancy.inter.fig}
  \end{figure}

Fig.~\ref{algo.redundancy.inter.fig} shows inter-monitor
redundancy when using Doubletree with $p=0.05$.

We first analyse the lower part of the graph.  The distribution of
hop counts for interfaces shows that most of the undiscovered
interfaces are far from the source.  As most of these nodes are
only visited by a single monitor (see
Fig.~\ref{redundancy.inter.global}), due to the nature of the
global stop set and the stop rule, the risk of missing them is
very high. Probably, with a higher value for $p$, we would have
elicited them. However, this solution would raise the redundancy
for destinations, as explained in Sec.~\ref{algo.tension.results}.
Those undiscovered nodes are, in a certain sense, the price to pay
to reduce the redundancy. Again, this fact represents the inherent
tension in the topology discovery problem. Some nodes are
sacrificed in order to reduce the redundancy.

If we compare Fig.~\ref{algo.redundancy.inter.fig} with
Fig.~\ref{redundancy.inter.global}, we can see that the redundancy
is strongly reduced.  The highest value for the median is only 6.
For the classical method, it is equal to the maximum, i.e., 24.
Furthermore, the highest quantiles between hop 4 and 13 are more
dissipated.

Finally, the right part of the graph, called \textit{all},
indicates that the median value is 2.  If we compare with
Fig.~\ref{redundancy.inter.global}, where the median equals 24, we
note that Doubletree allows a very strong reduction in
inter-monitor redundancy.

\section{Bloom Filters}\label{bf}

The algorithm presented in Sec.~\ref{algo.description} requires
that monitors exchange a set of
$(\mathrm{interface},\mathrm{destination})$ pairs.  The maximum
size of the global stop set will be the maximum number of
$(\mathrm{interface},\mathrm{destination})$ pairs in the data set
considered. Our study considers only 50,000 common destinations,
but skitter monitors probe towards a common set of on the order of
a million destinations.  Sharing a stop set based on this number
of destinations or even more could lead to a severe communication
overhead.  This should be avoided, or at least strongly limited,
in the case of a highly distributed measurement tool.

In order to reduce communication bandwidth cost, we propose to use
Bloom filters~\cite{bloom}, a technique that employs hash
functions to conduct lossy compression, and that has already seen
a number of networking applications, as described by Broder, A.
and Mitzenmacher in a 2002 survey~\cite{survey}.  A feature of
Bloom filters is that the tradeoffs between the degree of
compression they offer and the degree of error that their
lossiness introduces are well understood.

Bloom filters are used for verifying set membership.  The elements
of a set of data items, in this case the (interface,destination)
pairs of Doubletree's global stop set, are each hashed multiple
times to a vector, the filter. Subsequently, set membership can be
tested by examining the hash values that correspond to an
$(\mathrm{interface},\mathrm{destination})$ pair. If a pair is in
the set, the filter will always return true. However, there is a
finite, well defined, probability of a false positive for a pair
that is not in the set.

As we have described Doubletree in prior sections, each monitor
has full knowledge of what was discovered by the other monitors.
Each application of the stop set rule is thus taken with the
highest level of certainty. Now with the risk of false positives
from Bloom filters, some forwards probing along the tree-like
structure rooted at a destination will stop sooner than would
otherwise be the case.  The rate of false positives can be fine
tuned by adjusting such parameters as the size of vector employed,
and the number of hash functions. For a given number of elements
and a given vector size, for instance, an optimal number of hash
functions can be chosen to minimize the probability of false
positives.

We perform the same experiments as in Sec.~\ref{algo.tension} in
order to obtain a preliminary sense of the effect of Bloom filters
on Doubletree.  The methodology followed is the same as in
Sec.~\ref{algo.tension}, except for the stop set implementation.
We experiment with a low false positive rate. Choosing a vector
that contains ten times the number of bits as there are
$(\mathrm{interface},\mathrm{destination})$ pairs in the global
stop set, and using the optimal number of hash functions, five
(Fan et al.~\cite[Sec. 4.3]{bloomMath} make the same choices),
gives a false positive rate of 0.9\%.

Since a pair of IPv4 addresses consists of 64 bits, the Bloom
filter provides a 6.4:1 compression ratio.  This, of course, is a
first approximation, because the pair information could be
compressed using standard lossless compression techniques.
Likewise, Mitzenmacher~\cite{compressedBloom} has described
effective techniques for compression of Bloom filters; techniques
that have the effect of lowering the false positive rate. We have
yet to evaluate what the compression ratio would be if we were to
compare compressed pair lists with compressed Bloom filters.

  \begin{figure}[!tbp]
    \begin{center}
      \includegraphics[width=8cm]{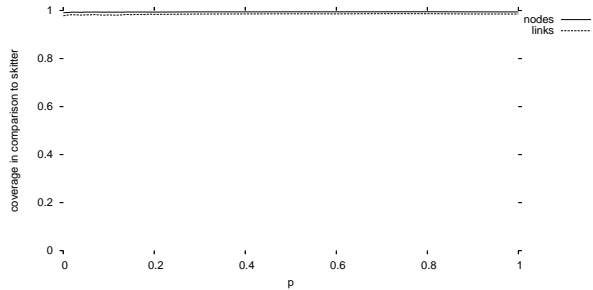}
    \end{center}
    \caption{Links and nodes coverage while using Bloom filters}
    \label{bf.coverage}
  \end{figure}

Fig.~\ref{bf.coverage} presents the coverage in terms of links and
nodes when using a Bloom filter with the parameters just
described.  The level of coverage is runs from 99\% ($p=0$) to
99.5\% for nodes and from 97\% ($p=0$) to 98,6\% for links. The
coverage values are more uniform across values of $p$ than they
were for the standard global stop set.  This coverage is better
than the results we describe in
Sec.~\ref{algo.tension.methodology}, that do not use Bloom
filters.

This result is counterintuitive because we would expect that false
positives would have the effect of constraining exploration rather
than promoting it. It seems all the more surprising that coverage
should increase at $p=1$, when the global stop set, or its Bloom
filter replacement, is not even applied for backwards probing. To
address this second question first, recall that when an interface
at hop distance $h$ does not respond to probes, $h$ is halved and
halved again until a responding interface is found. Both forwards
and backwards probing then takes place from this new $h$.  In the
data set that we consider, approximately 40\% of destinations do
not respond. Consequently, when $p=1$, in fully 40\% of the cases
Doubletree performs forwards probing, and can use the global stop
set implemented as a Bloom filter.

Regarding the increase in coverage overall, we can only speculate
as to the reasons.  It might be that simply the fact of
introducing a degree of randomness into the exploration process
enhances discovery.  Perhaps explorations that are blocked by a
false positive leave the way open for further explorations because
fewer pairs enter the stop set.  An assessment of the effect of
introducing randomness into the stopping rule (both false
positives and false negatives, independently of the use of Bloom
filters for the stop set) is a subject for future work.

These initial results are encouraging from the point of view of
node and link coverage.  However, the price comes in the form of
additional probe traffic. The reduction in the number of probes
compared to the classical approach is only 8\% when $p=0$.
Nevertheless, the number of probes sent decreases as $p$
increases. It oscillates between 8\% and 59\%.  The gross
redundancy on internal interfaces is also higher when $p=0$ (398)
but it also decreases when $p$ increases. One concern is that the
gains are reversed regarding inter-monitor redundancy on
destinations.  The \nth{95} percentile redundancy is at the
maximum (i.e., 24) for each value of $p$.  A full exploration of
the trade-offs involved in the use of Bloom filters is a subject
for future work.

\section{Conclusion}\label{conclusion}

In this paper, we quantify the amount of redundancy in classical
internet topology discovery approaches by taking into account the
perspective from the single monitor (intra-monitor) and that of
the entire system (inter-monitor).  In the intra-monitor case, we
find that interfaces close to the monitor suffer from a high
number of repeat visits. Concerning inter-monitor redundancy, we
see that a large portion of interfaces are visited by all
monitors.

In order to scale up classical approaches, we have proposed
Doubletree, an algorithm that significantly reduces the
duplication of effort while discovering nearly the same set of
nodes and links. Doubletree simultaneously meets the conflicting
demands of reducing intra- and inter-monitor redundancy. We
describe how to tune a single parameter for Doubletree in order to
obtain an acceptable trade-off between redundancy and coverage.

Doubletree introduces communication between monitors.  To address
the problem of bandwidth consumption, we propose to encode this
communication through the use of Bloom filters. Surprisingly, we
find that this encoding technique, though it generates false
positives that might seem to constrain exploration, can actually
increase the coverage of nodes and links.

For future work, we plan to study in detail the trade-offs
involved in the use Bloom filters. How does the choice of vector
size and number of hash functions affect levels of redundancy and
coverage?  We will also evaluate the relevance of introducing a
certain level of false negatives to the stop set.

A probing technique that starts probing at a hop $h$ far from the
monitor has a non zero probability $p$ of hitting a destination
with its first probe. This has serious consequences when scaling
up the number of monitors.  Indeed, the average impact on
destinations will grow linearly as a function of $m$, the number
of monitors.  As $m$ increases, the risk that probing will appear
to be a DDoS attack will grow.

In order to permit greater scaling, we have started to investigate
techniques for dividing up the monitor set and the destination set
into subsets that we call \dfn{clusters}. By placing an upper
bound on the number of monitors in a cluster, we hope to place a
definitive upper bound on inter-monitor redundancy for destination
interfaces.  Clustering will have effects on redundancy and
coverage, and we are investigating these trade-offs.

\section*{Acknowledgments}

Without the data provided by k claffy and her team at
\textsc{Caida}, this research would not have been possible.  They
have also provided many helpful comments; for this, we thank Andre
Broido in particular.  Pierre Lafon and his team at the Centre de
Calcul Formel Médicis, Laboratoire Stix, Ecole Polytechnique,
kindly gave us access to their computing cluster, allowing faster
and easier simulations.  We thank our partners in the
traceroute@home project, notably Matthieu Latapy, Alessandro
Vespignani, and Alain Barrat, for their support and feedback.

\bibliographystyle{IEEE}
\bibliography{Bibliography}

\clearpage

\appendix

\onecolumn

\section{Intra-monitor redundancy plots}\label{appendix.redundancy}

This appendix presents the intra-monitor redundancy plots for all
24 skitter monitors that form the basis for the study in this
paper.  For each monitor, we show the redundancy of a classic
topology discovery system.  For 18 of the monitors, we also show
the result of a system applying the Doubletree algorithm with
parameter $p=0.05$.  In each of these cases, the monitor is the
last of the 24 to conduct its probing, using the global stop set
that has been passed to it by the other monitors.

\begin{figure*}[htbp]
  \begin{center}
    \subfigure[\texttt{apan-jp} classic]{\label{appendix.redundancy.intra.apan.skitter}
        \includegraphics[width=5cm]{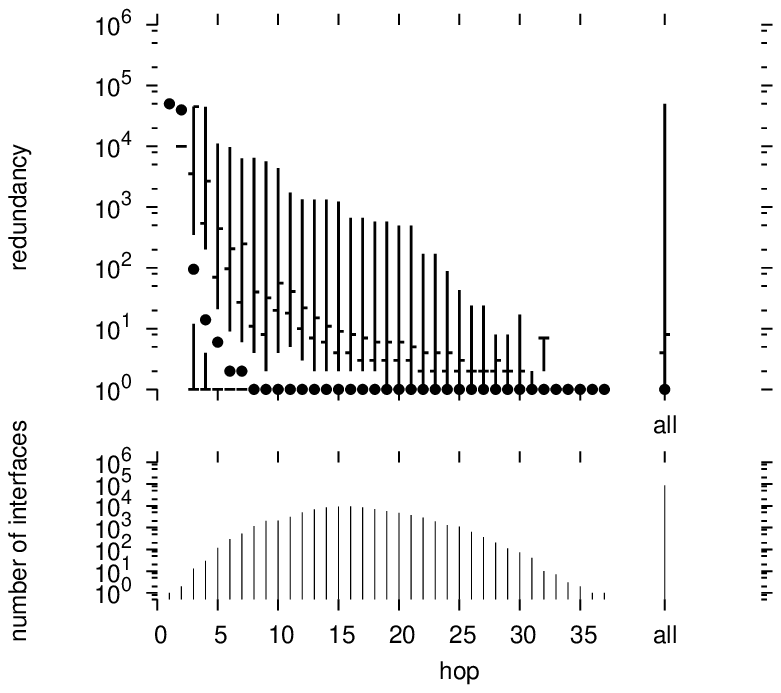}}
    \subfigure[\texttt{cam} classic]{\label{appendix.redundancy.intra.cam.classic}
        \includegraphics[width=5cm]{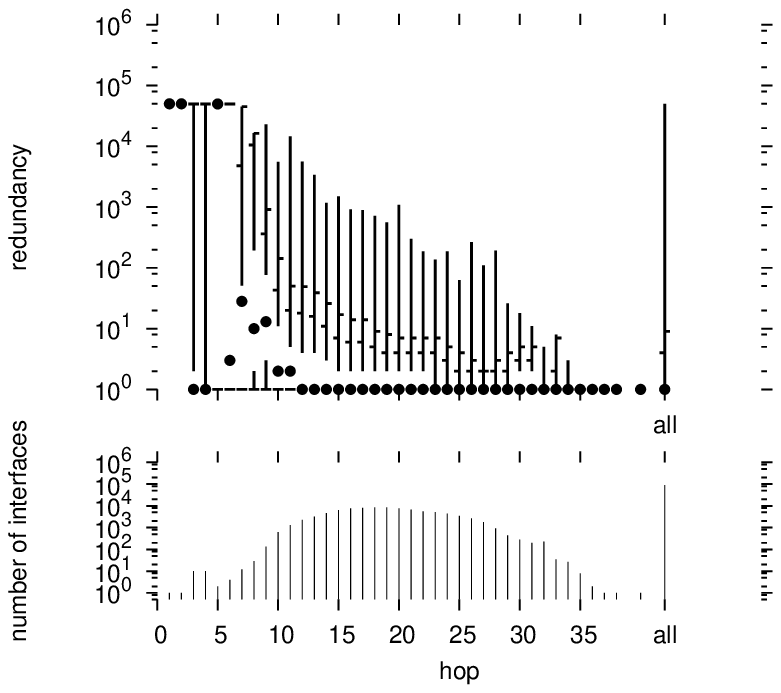}}
    \subfigure[\texttt{h-root} classic]{\label{appendix.redundancy.intra.hroot.classic}
        \includegraphics[width=5cm]{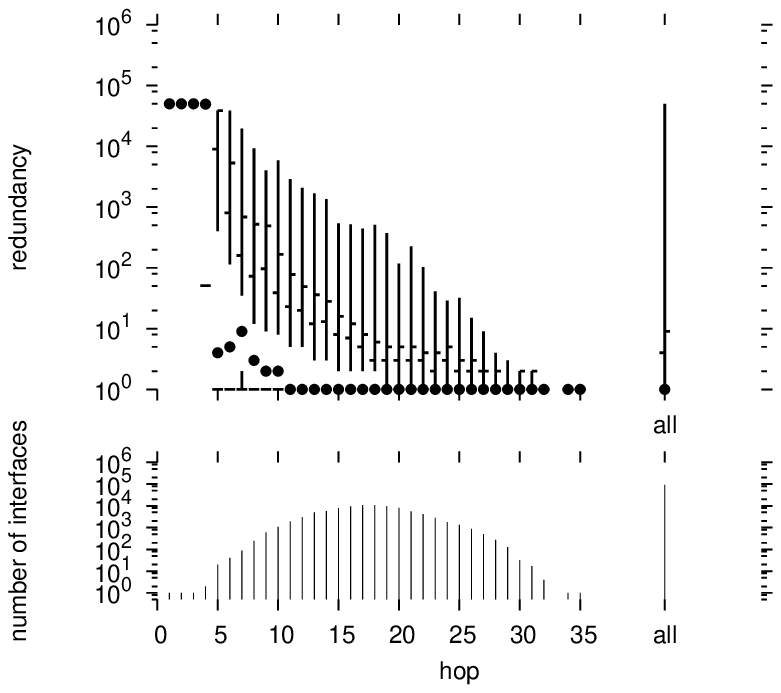}}
    \subfigure[\texttt{i-root} classic]{\label{appendix.redundancy.intra.iroot.classic}
        \includegraphics[width=5cm]{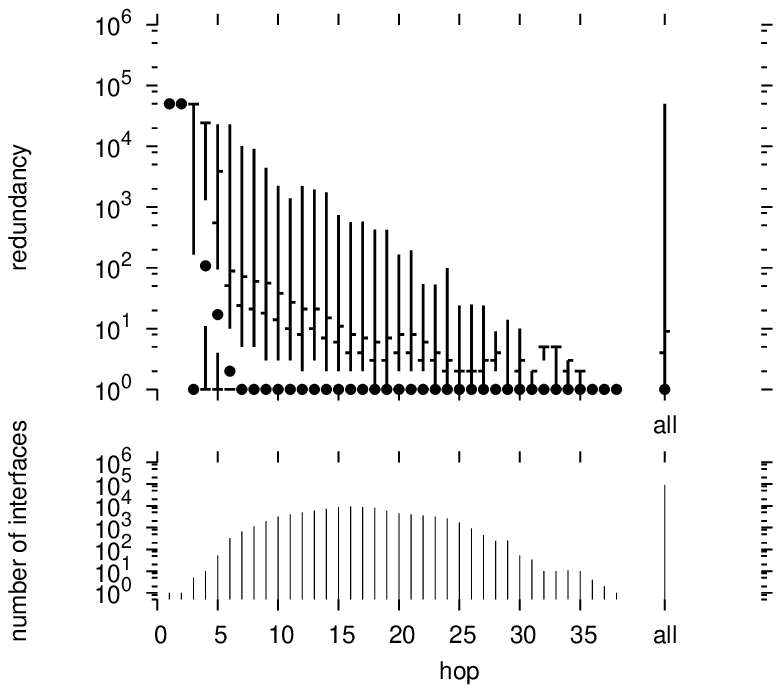}}
    \subfigure[\texttt{k-root} classic]{\label{appendix.redundancy.intra.kroot.classic}
        \includegraphics[width=5cm]{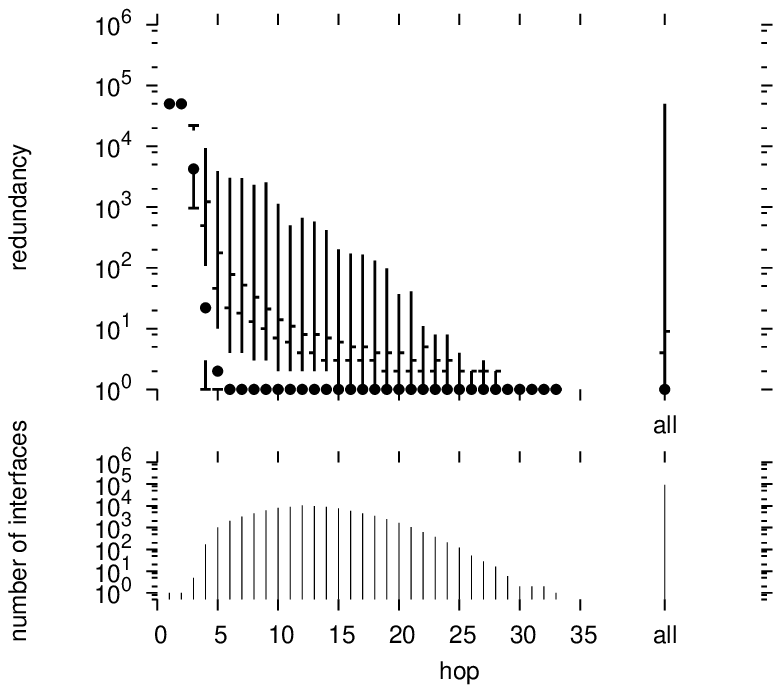}}
    \subfigure[\texttt{uoregon} classic]{\label{appendix.redundancy.intra.uoregon.classic}
        \includegraphics[width=5cm]{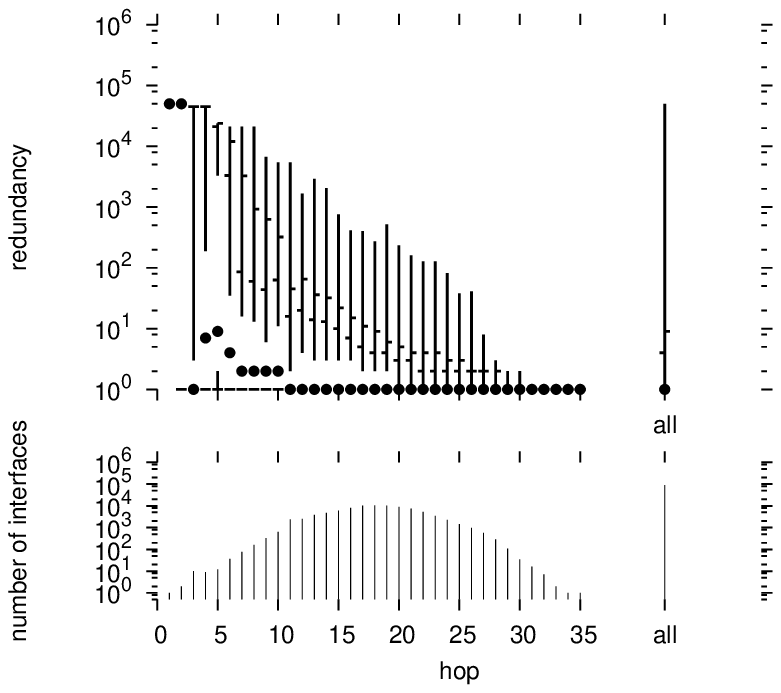}}
  \end{center}
  \caption{\texttt{apan-jp}, \texttt{cam}, \texttt{h-root}, \texttt{i-root}, \texttt{k-root}, and \texttt{uoregon}}
\end{figure*}

\clearpage

\begin{figure*}[htbp]
  \begin{center}
    \mbox{
        \subfigure[\texttt{arin} classic]{\label{appendix.redundancy.intra.arin.classic}
            \includegraphics[width=5cm]{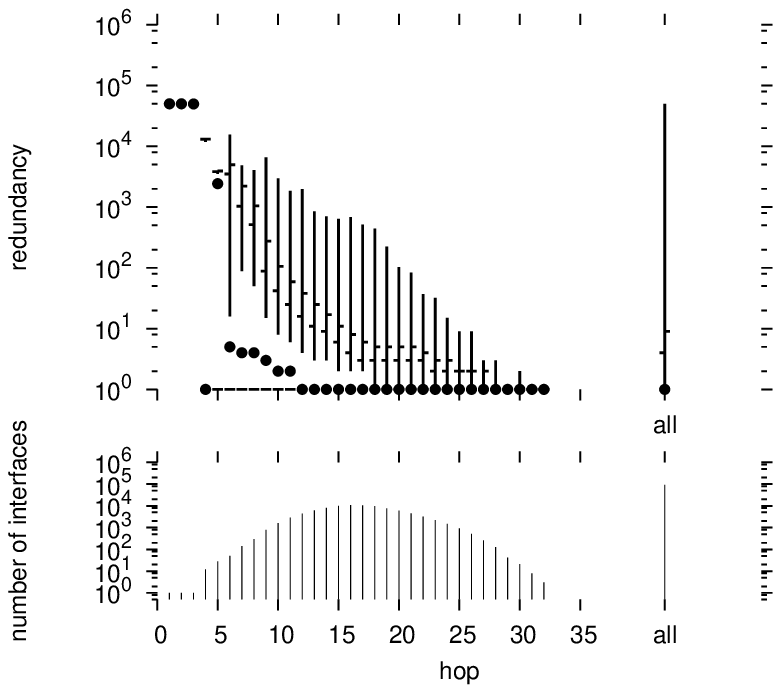}}
            \qquad
        \subfigure[\texttt{arin} Doubletree ($p=0.05$)]{\label{appendix.redundancy.intra.arin.doubletree.p005}
            \includegraphics[width=5cm]{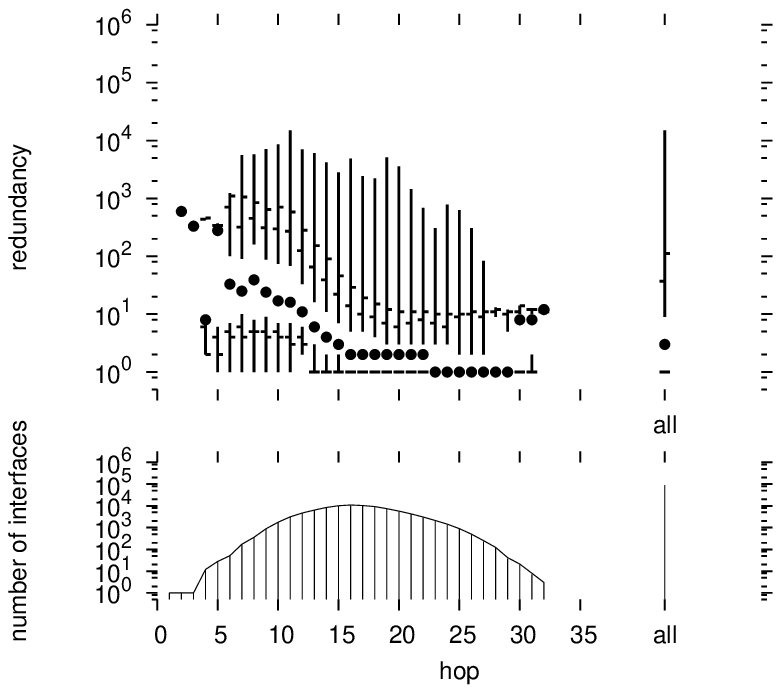}}
    }
    \mbox{
        \subfigure[\texttt{b-root} classic]{\label{appendix.redundancy.intra.broot.classic}
            \includegraphics[width=5cm]{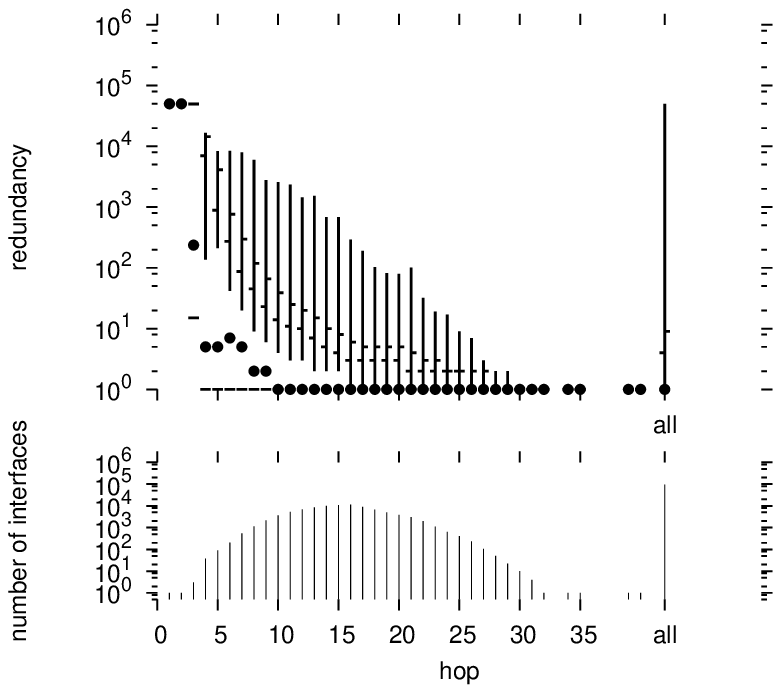}}
            \qquad
        \subfigure[\texttt{b-root} Doubletree ($p=0.05$)]{\label{appendix.redundancy.intra.broot.doubletree.p005}
            \includegraphics[width=5cm]{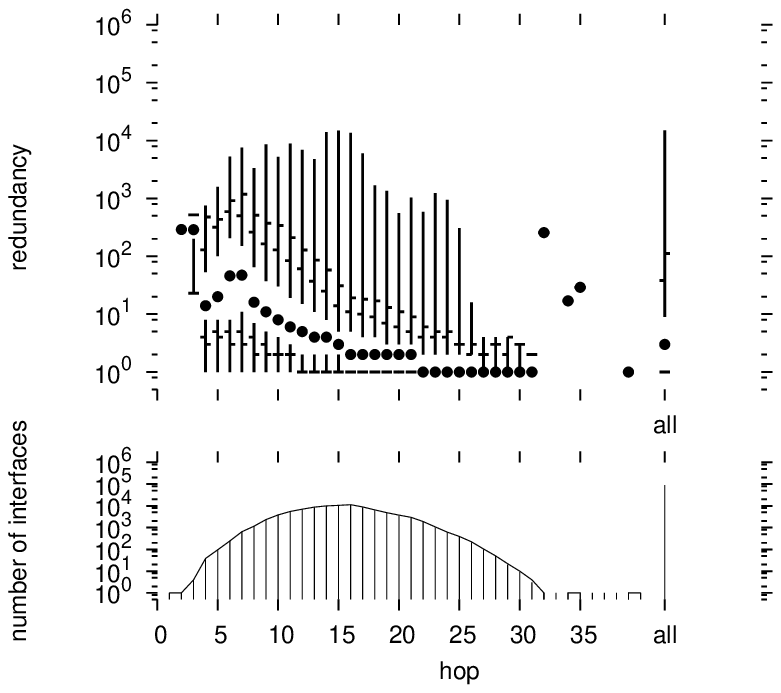}}
    }
    \mbox{
        \subfigure[\texttt{cdg-rssac} classic]{\label{appendix.redundancy.intra.cdg.classic}
            \includegraphics[width=5cm]{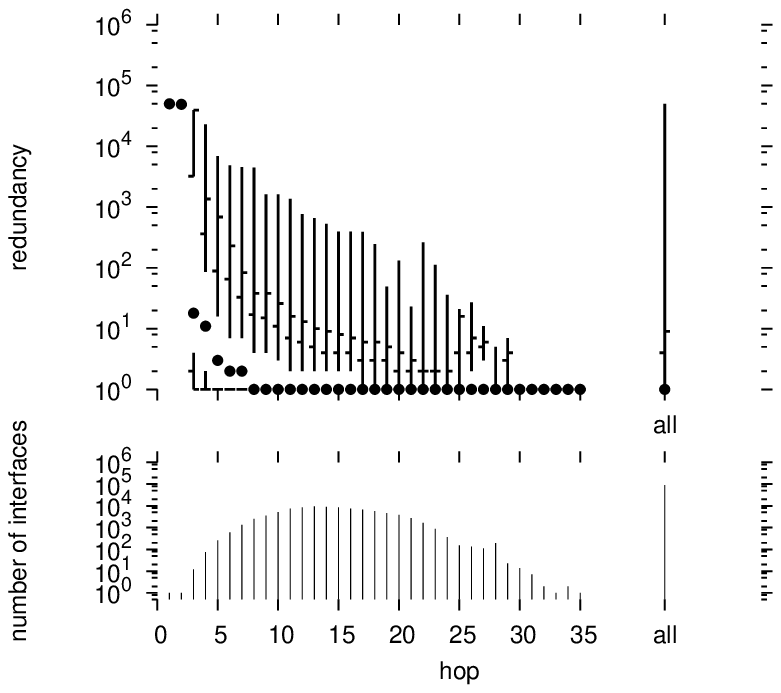}}
            \qquad
        \subfigure[\texttt{cdg-rssac} Doubletree ($p=0.05$)]{\label{appendix.algo.intra.cdg.doubletree.p005}
            \includegraphics[width=5cm]{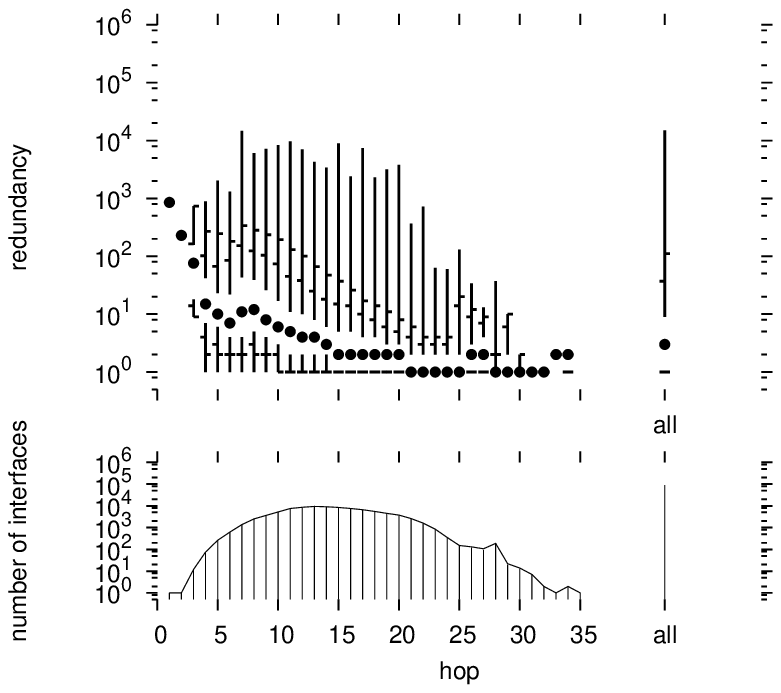}}
    }
  \end{center}
  \caption{\texttt{arin}, \texttt{b-root}, and \texttt{cdg-rssac}}
\end{figure*}

\clearpage

\begin{figure*}[htbp]
  \begin{center}
    \mbox{
        \subfigure[\texttt{champagne} classic]{\label{appendix.redundancy.intra.champ.classic}
            \includegraphics[width=5cm]{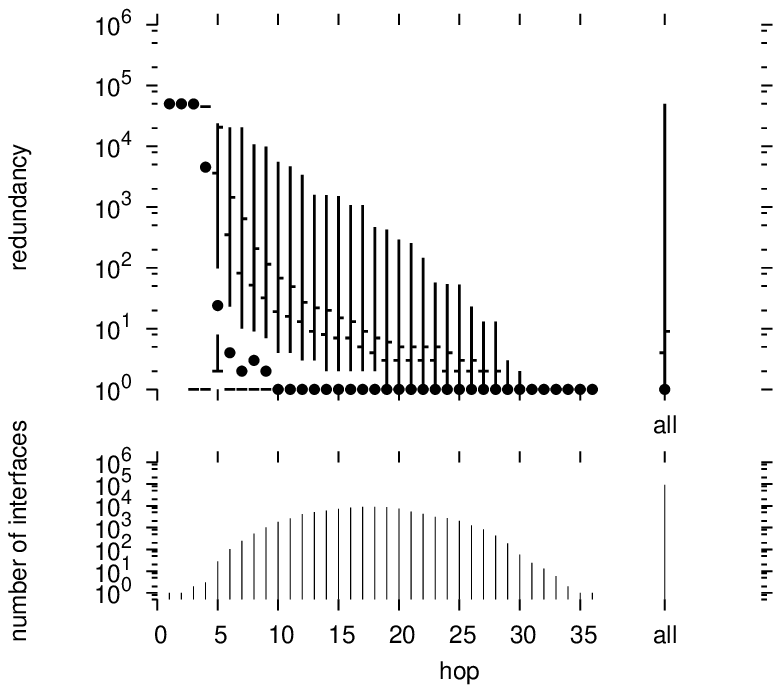}}
            \qquad
        \subfigure[\texttt{champagne} Doubletree ($p=0.05$)]{\label{appendix.algo.intra.champ.doubletree.p005}
            \includegraphics[width=5cm]{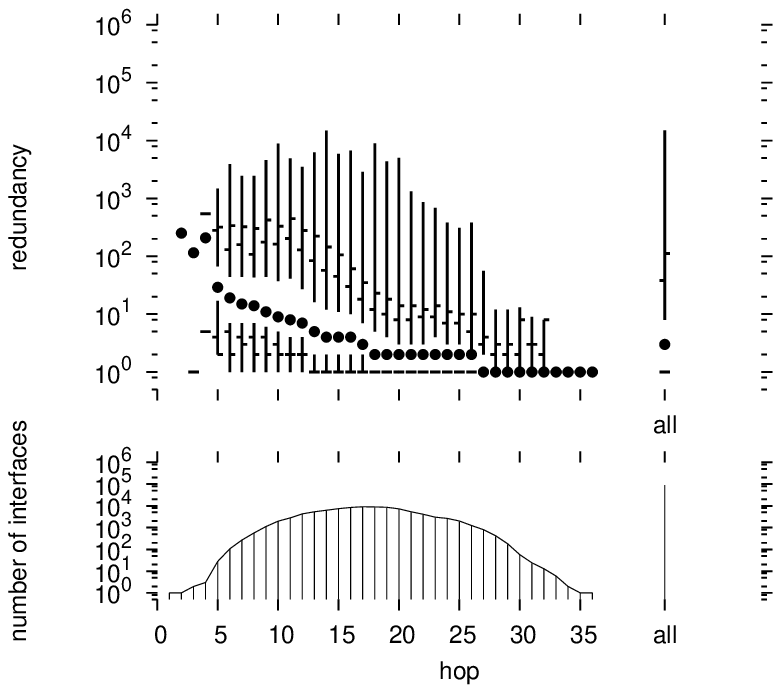}}
    }
    \mbox{
        \subfigure[\texttt{d-root} classic]{\label{appendix.redundancy.intra.droot.classic}
            \includegraphics[width=5cm]{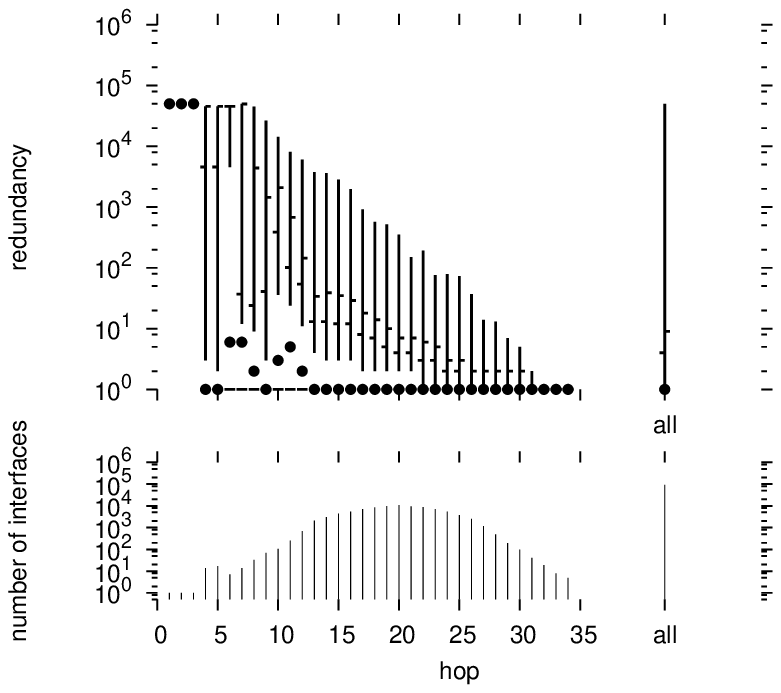}}
            \qquad
        \subfigure[\texttt{d-root} Doubletree ($p=0.05$)]{\label{appendix.algo.intra.droot.doubletree.p005}
            \includegraphics[width=5cm]{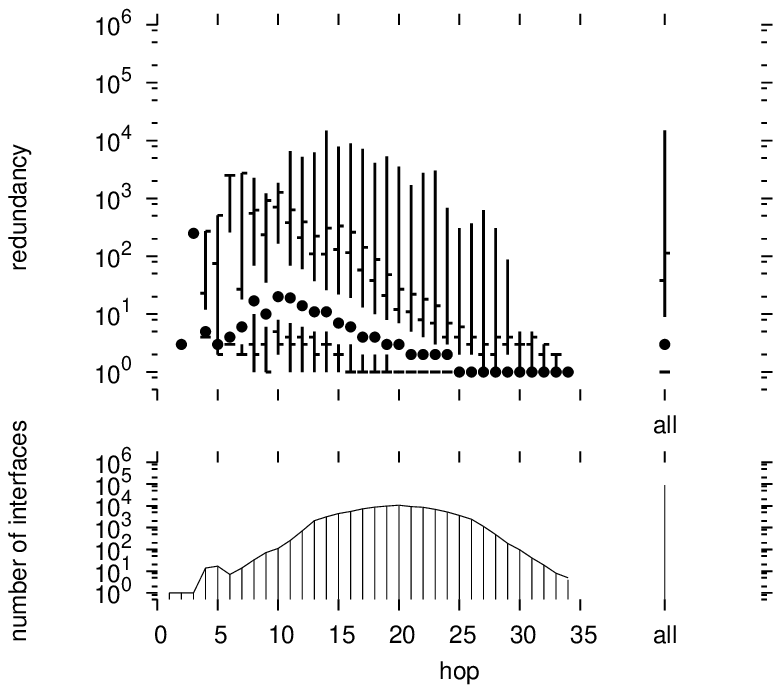}}
    }
    \mbox{
        \subfigure[\texttt{e-root} classic]{\label{appendix.redundancy.intra.eroot.classic}
            \includegraphics[width=5cm]{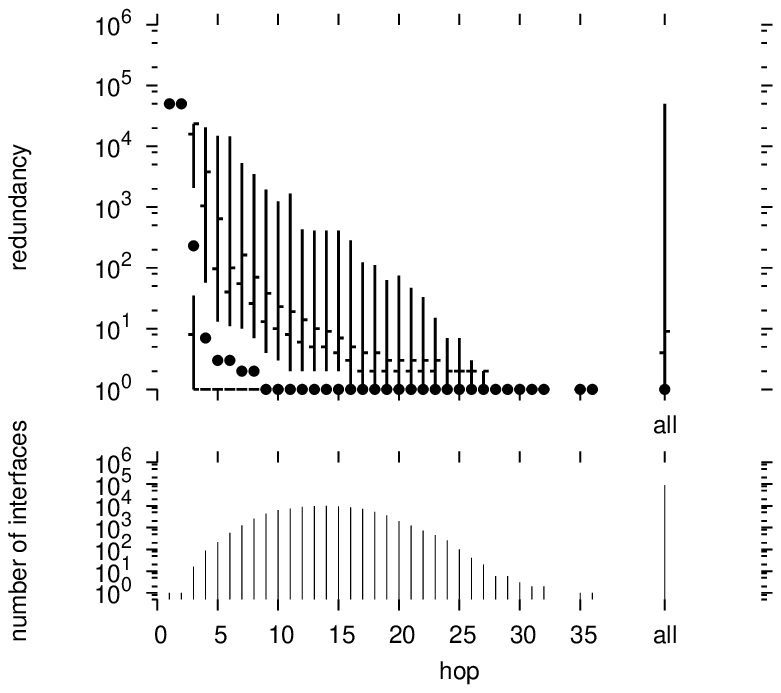}}
            \qquad
        \subfigure[\texttt{e-root} Doubletree ($p=0.05$)]{\label{appendix.algo.intra.eroot.doubletree.p005}
            \includegraphics[width=5cm]{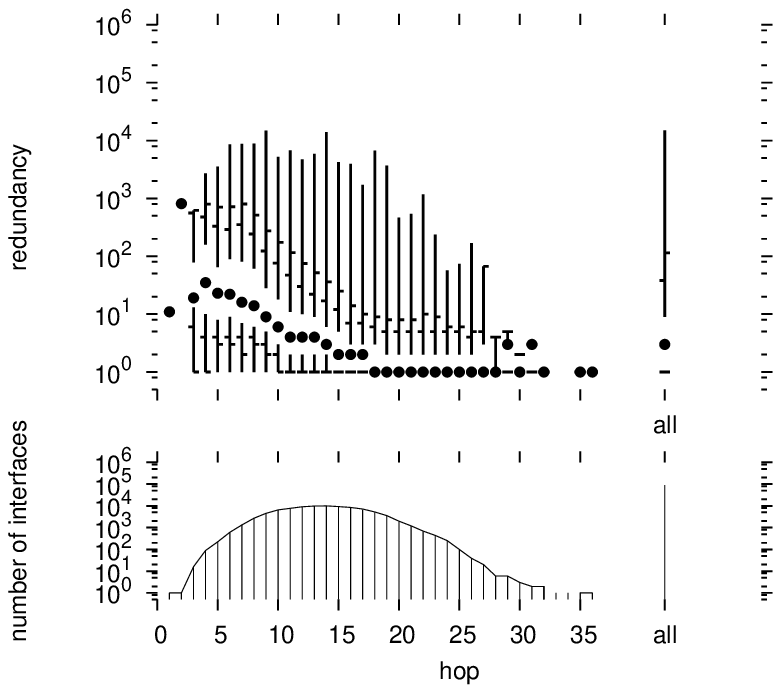}}
    }
  \end{center}
  \caption{\texttt{champagne}, \texttt{d-root}, and \texttt{e-root}}
\end{figure*}

\clearpage

\begin{figure*}[htbp]
  \begin{center}
    \mbox{
        \subfigure[\texttt{f-root} classic]{\label{appendix.redundancy.intra.froot.classic}
            \includegraphics[width=5cm]{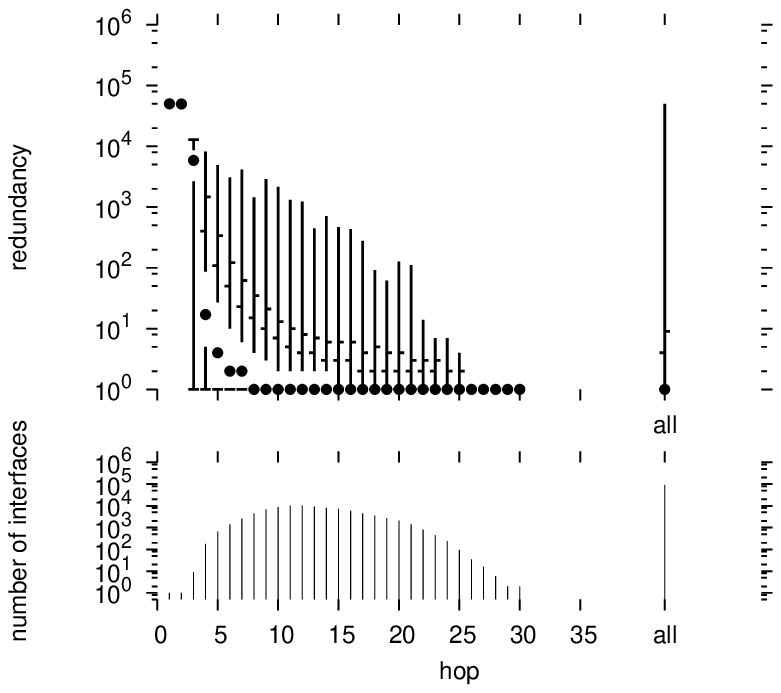}}
            \qquad
        \subfigure[\texttt{f-root} Doubletree ($p=0.05$)]{\label{appendix.algo.intra.froot.classic.doubletree.p005}
            \includegraphics[width=5cm]{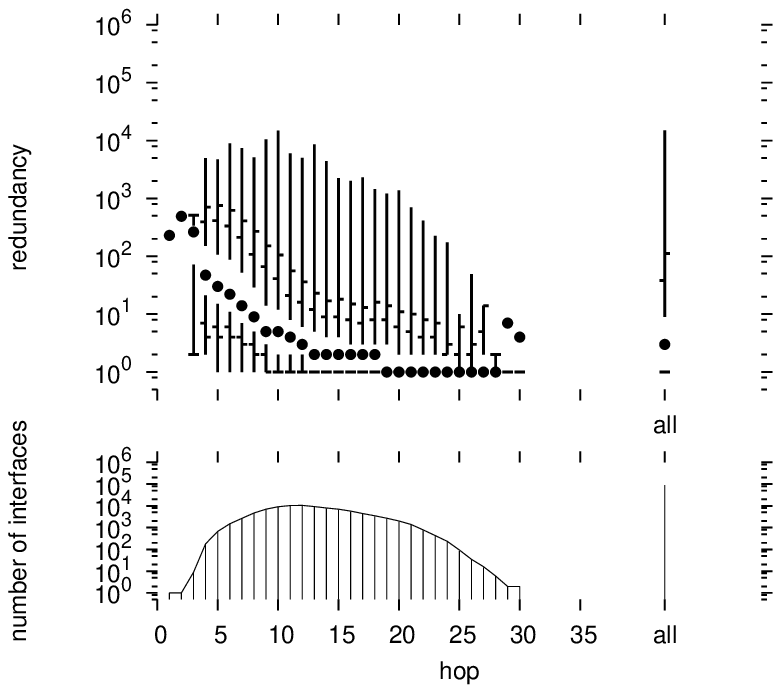}}
    }
    \mbox{
        \subfigure[\texttt{g-root} classic]{\label{appendix.redundancy.intra.groot.classic}
            \includegraphics[width=5cm]{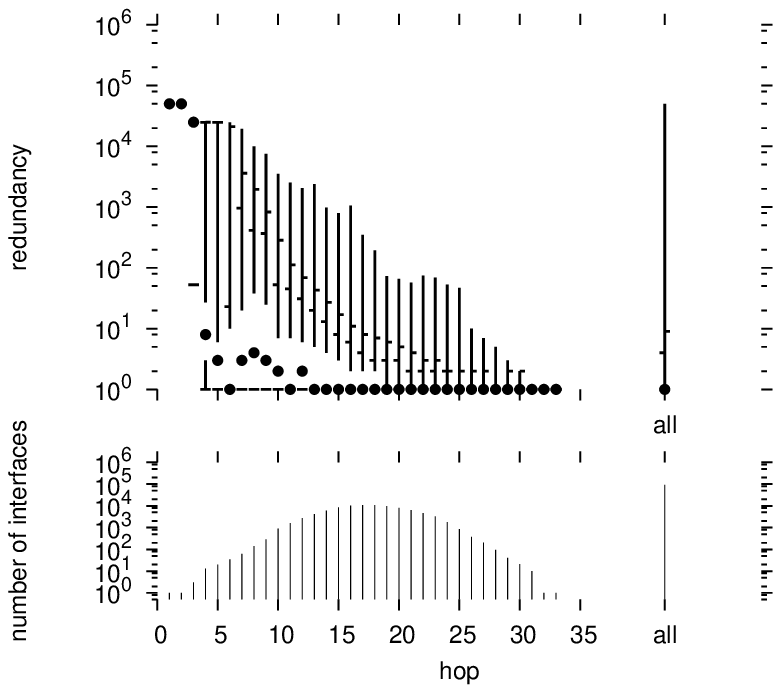}}
            \qquad
        \subfigure[\texttt{g-root} Doubletree ($p=0.05$)]{\label{appendix.algo.intra.groot.doubletree.p005}
            \includegraphics[width=5cm]{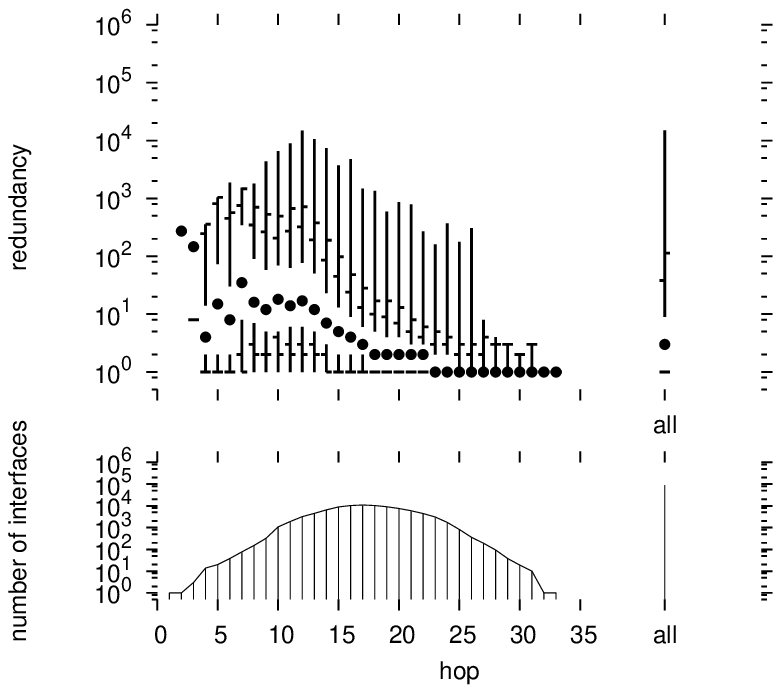}}
    }
    \mbox{
        \subfigure[\texttt{iad} classic]{\label{appendix.redundancy.intra.iad.classic}
            \includegraphics[width=5cm]{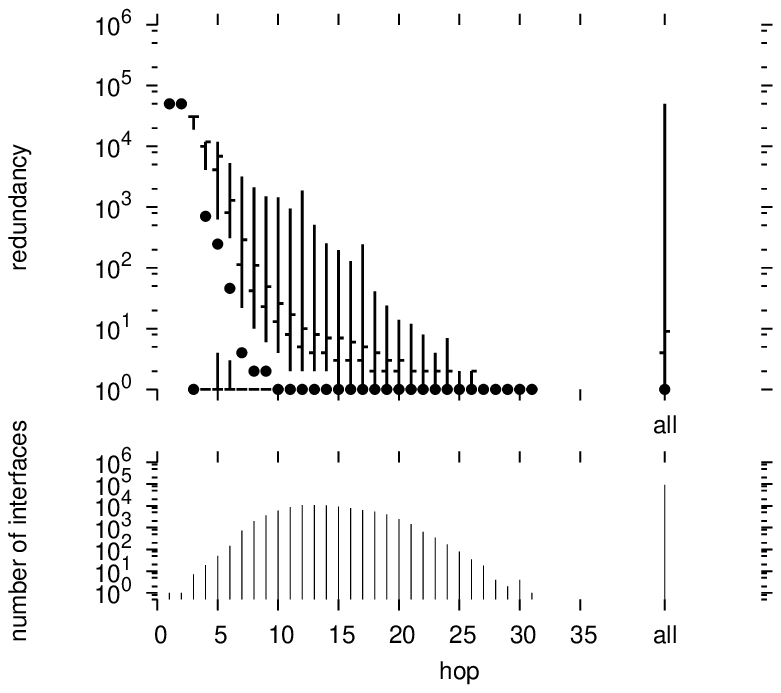}}
            \qquad
        \subfigure[\texttt{iad} Doubletree ($p=0.05$)]{\label{appendix.algo.intra.iad.doubletree.p005}
            \includegraphics[width=5cm]{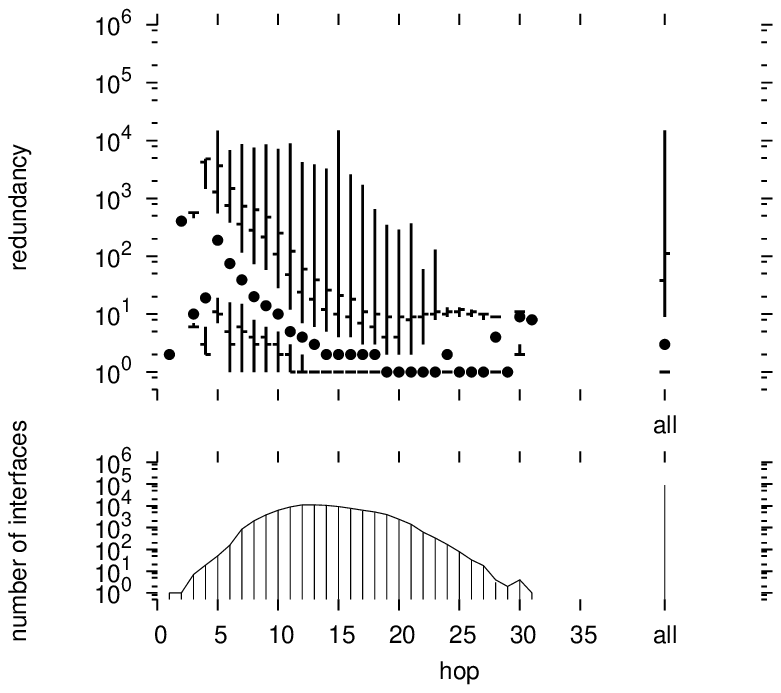}}
    }
  \end{center}
  \caption{\texttt{f-root}, \texttt{g-root}, and \texttt{iad}}
\end{figure*}

\clearpage

\begin{figure*}[htbp]
  \begin{center}
    \mbox{
    \subfigure[\texttt{ihug} classic]{\label{appendix.redundancy.intra.ihug.classic}
      \includegraphics[width=5cm]{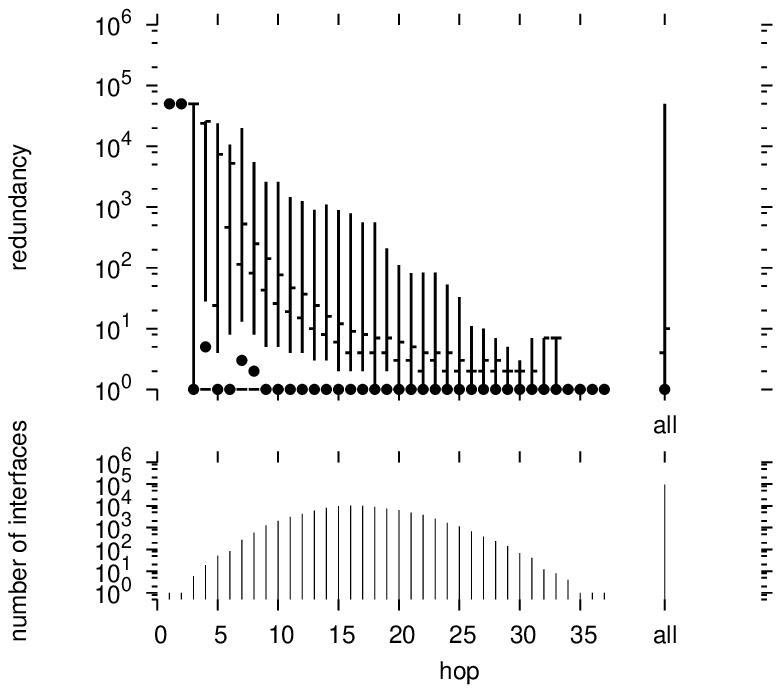}}
      \qquad
    \subfigure[\texttt{ihug} Doubletree ($p=0.05$)]{\label{appendix.algo.intra.ihug.doubletree.p005}
      \includegraphics[width=5cm]{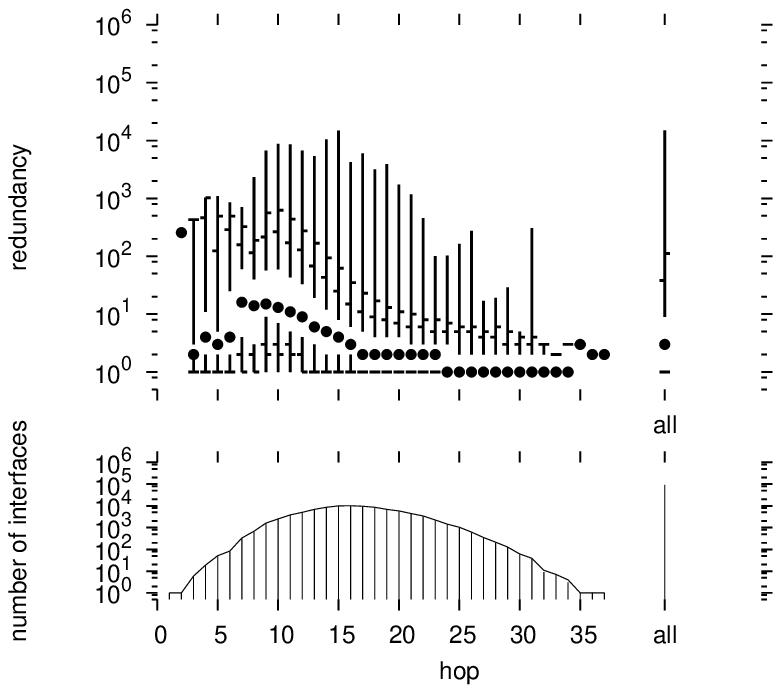}}
    }
    \mbox{
    \subfigure[\texttt{k-peer} classic]{\label{appendix.redundancy.intra.kpeer.classic}
      \includegraphics[width=5cm]{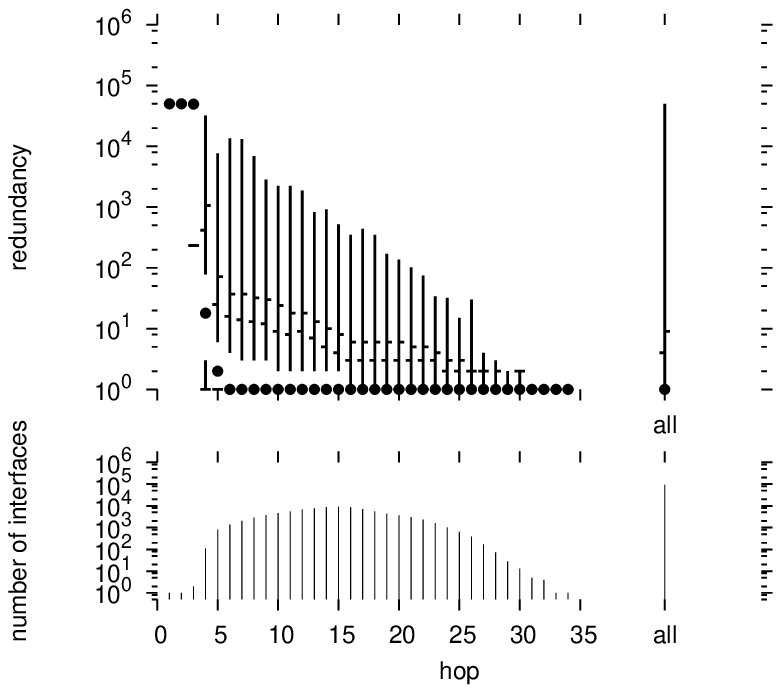}}
      \qquad
    \subfigure[\texttt{k-peer} Doubletree ($p=0.05$)]{\label{appendix.algo.intra.kpeer.doubletree.p005}
      \includegraphics[width=5cm]{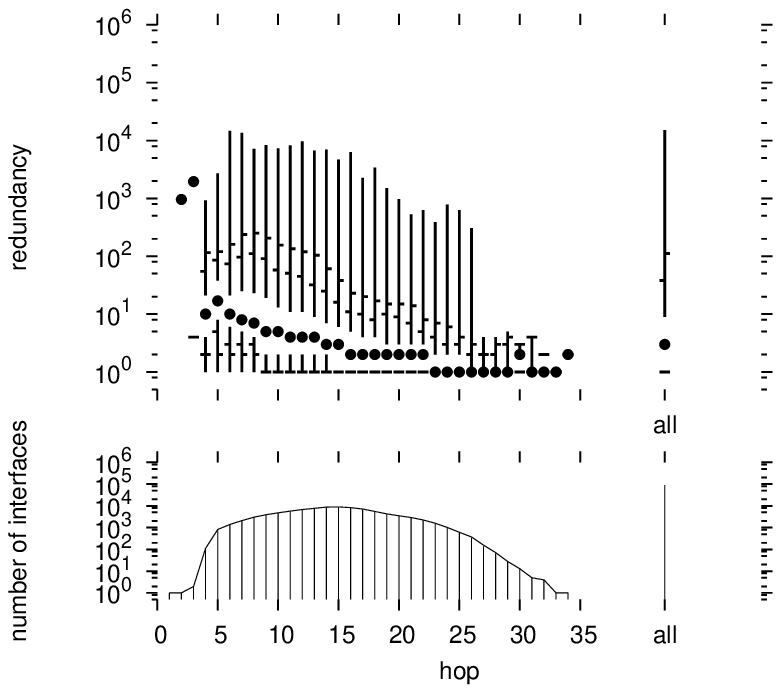}}
    }
    \mbox{
    \subfigure[\texttt{lhr} classic]{\label{appendix.redundancy.intra.lhr.classic}
      \includegraphics[width=5cm]{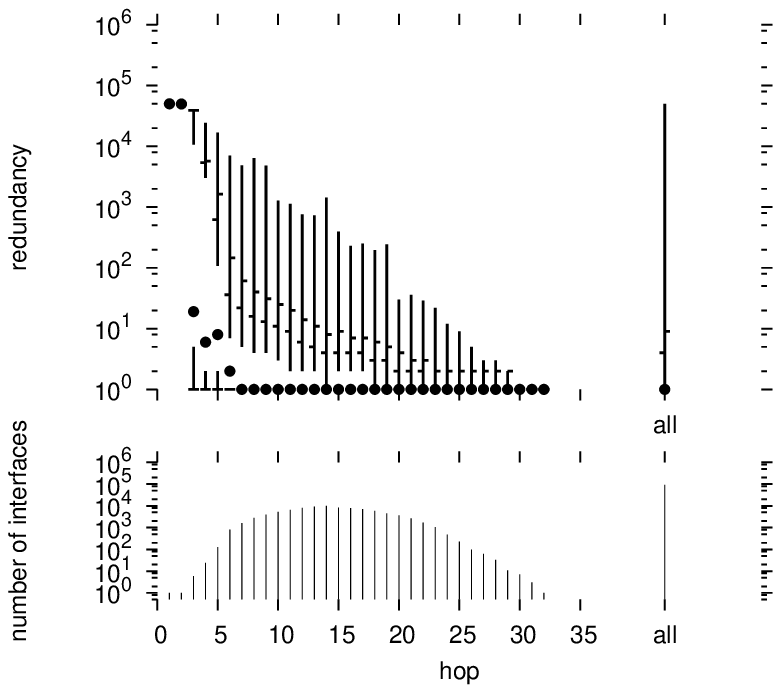}}
      \qquad
    \subfigure[\texttt{lhr} Doubletree ($p=0.05$)]{\label{appendix.algo.intra.lhr.doubletree.p005}
      \includegraphics[width=5cm]{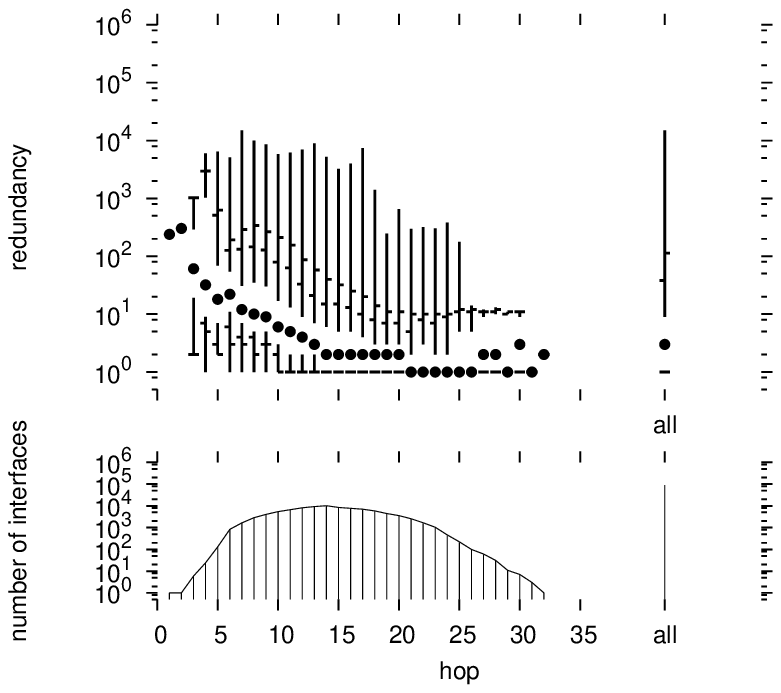}}
    }
  \end{center}
  \caption{\texttt{ihug}, \texttt{k-peer}, and \texttt{lhr}}
\end{figure*}

\clearpage

\begin{figure*}[htbp]
  \begin{center}
    \mbox{
        \subfigure[\texttt{m-root} classic]{\label{appendix.redundancy.intra.mroot.classic}
            \includegraphics[width=5cm]{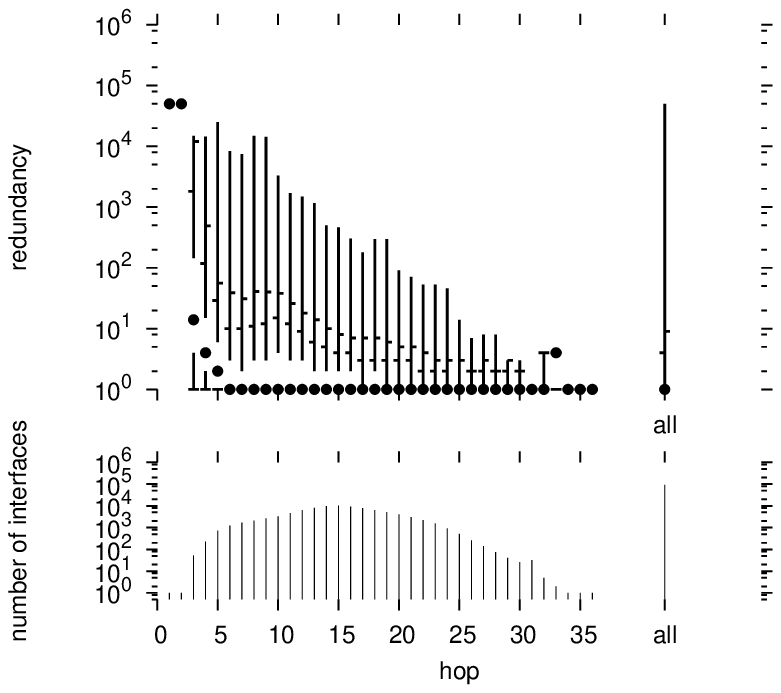}}
        \qquad
        \subfigure[\texttt{m-root} Doubletree ($p=0.05$)]{\label{appendix.algo.intra.mroot.doubletree.p005}
            \includegraphics[width=5cm]{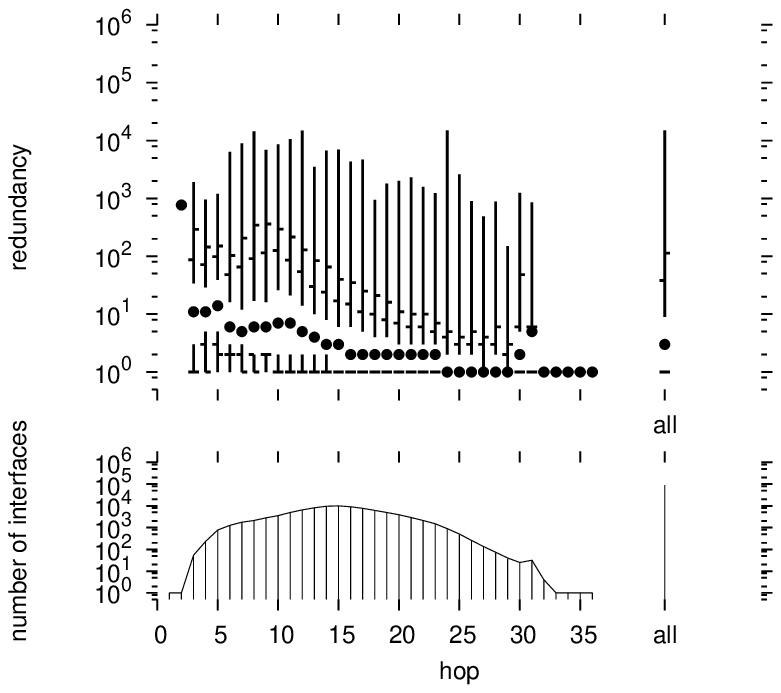}}
    }
    \mbox{
        \subfigure[\texttt{mwest} classic]{\label{appendix.redundancy.intra.mwest.classic}
            \includegraphics[width=5cm]{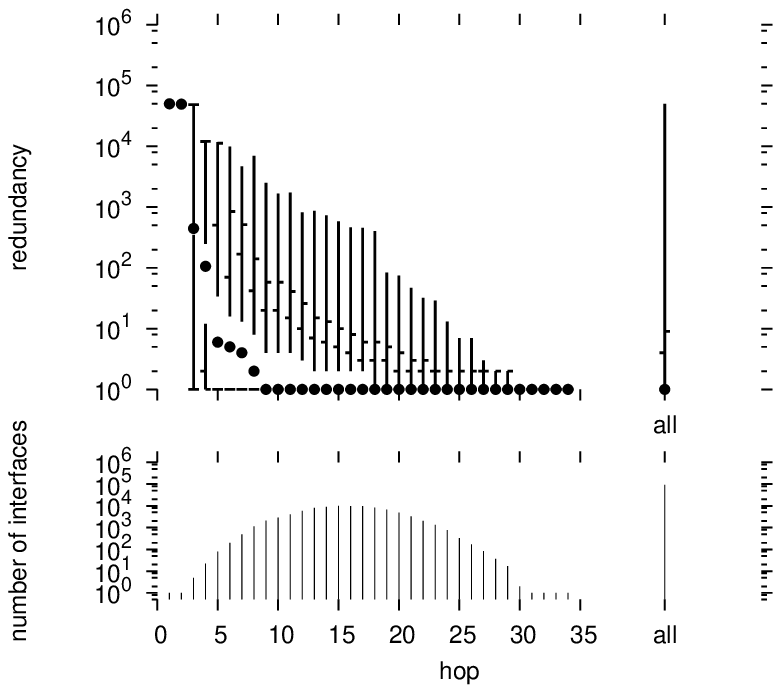}}
        \qquad
        \subfigure[\texttt{mwest} Doubletree ($p=0.05$)]{\label{appendix.algo.intra.mwest.doubletree.p005}
            \includegraphics[width=5cm]{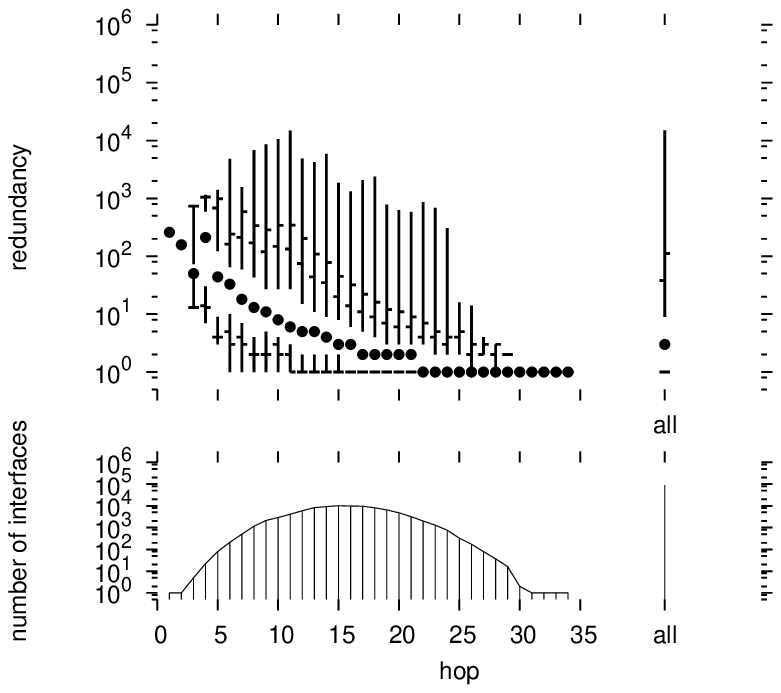}}
    }
    \mbox {
        \subfigure[\texttt{nrt} classic]{\label{appendix.redundancy.intra.nrt.classic}
            \includegraphics[width=5cm]{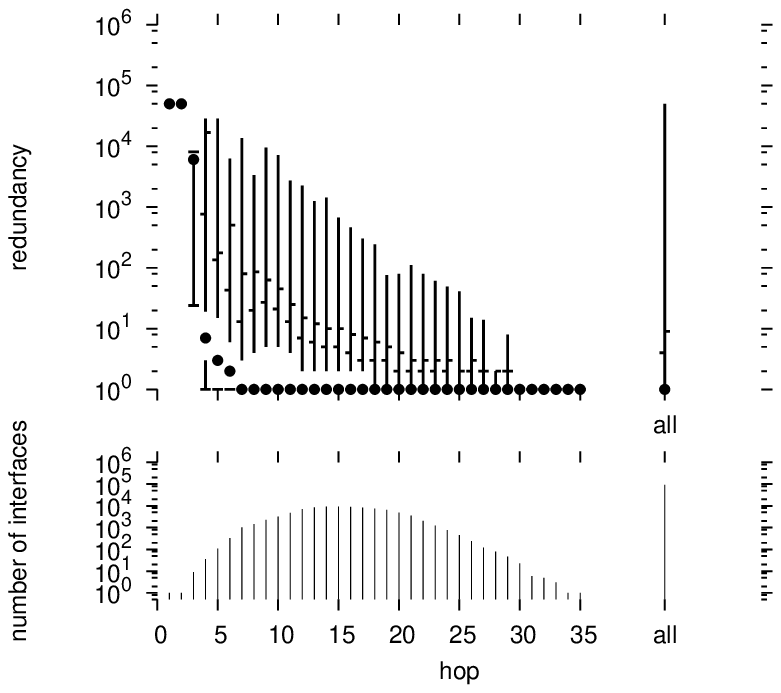}}
        \qquad
        \subfigure[\texttt{nrt} Doubletree ($p=0.05$)]{\label{appendix.algo.intra.nrt.doubletree.p005}
            \includegraphics[width=5cm]{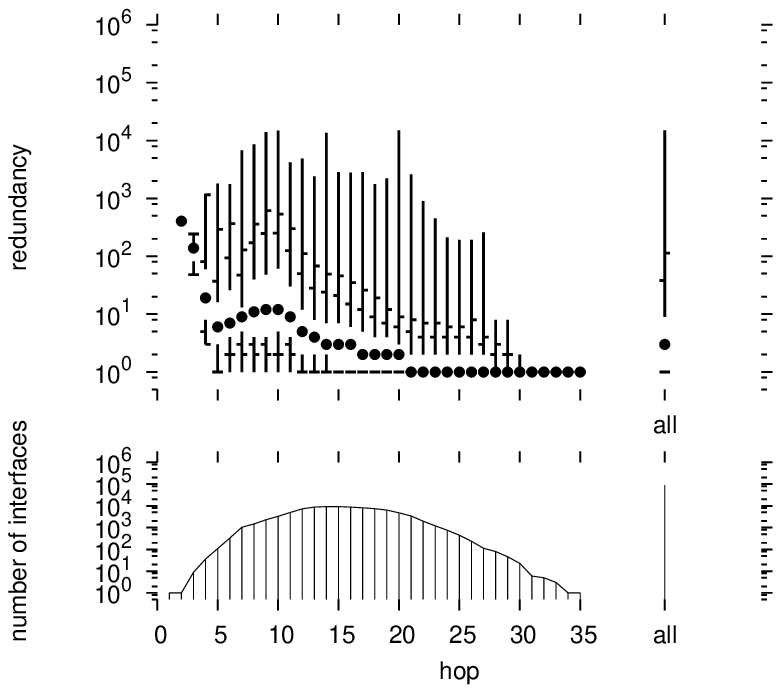}}
    }
  \end{center}
  \caption{\texttt{m-root}, \texttt{mwest}, and \texttt{nrt}}
\end{figure*}

\clearpage

\begin{figure*}[htbp]
  \begin{center}
    \mbox{
        \subfigure[\texttt{riesling} classic]{\label{appendix.redundancy.intra.riesling.classic}
            \includegraphics[width=5cm]{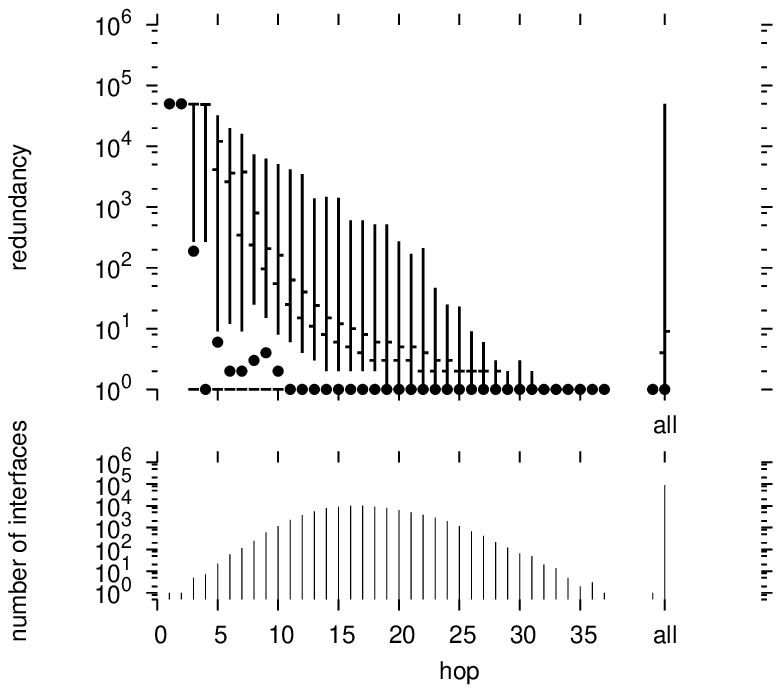}}
            \qquad
        \subfigure[\texttt{riesling} Doubletree ($p=0.05$)]{\label{appendix.algo.intra.riesling.doubletree.p005}
            \includegraphics[width=5cm]{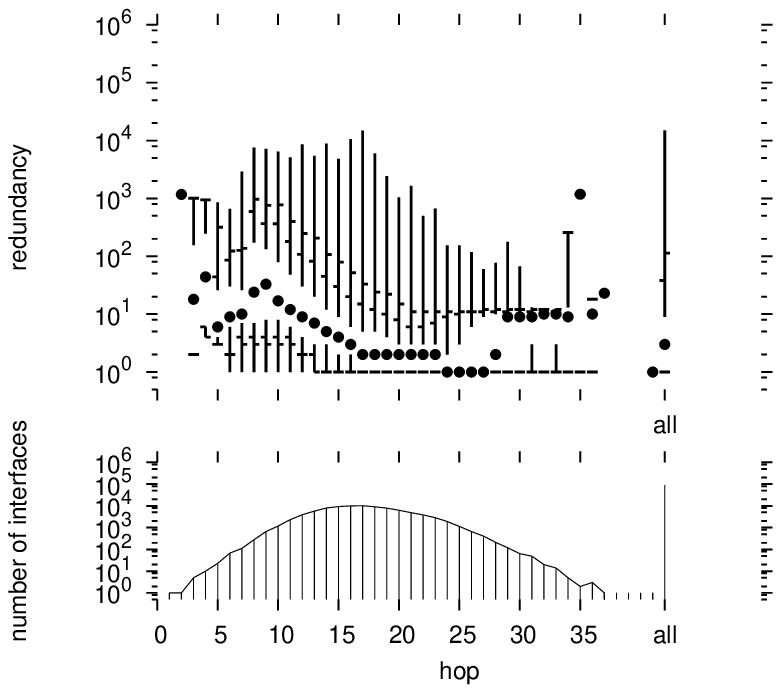}}
    }
    \mbox{
        \subfigure[\texttt{sjc} classic]{\label{appendix.redundancy.intra.sjc.classic}
            \includegraphics[width=5cm]{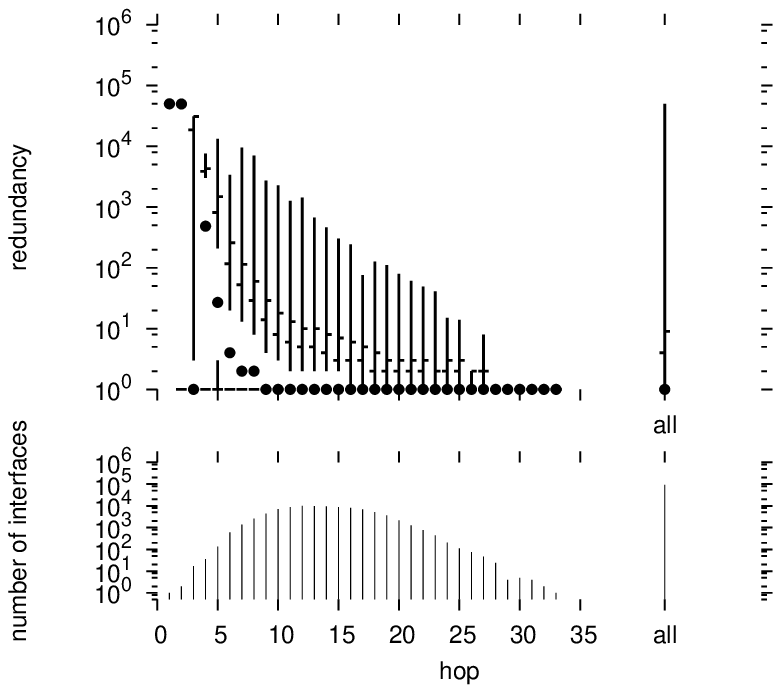}}
            \qquad
        \subfigure[\texttt{sjc} Doubletree ($p=0.05$)]{\label{appendix.algo.intra.sjc.doubletree.p005}
            \includegraphics[width=5cm]{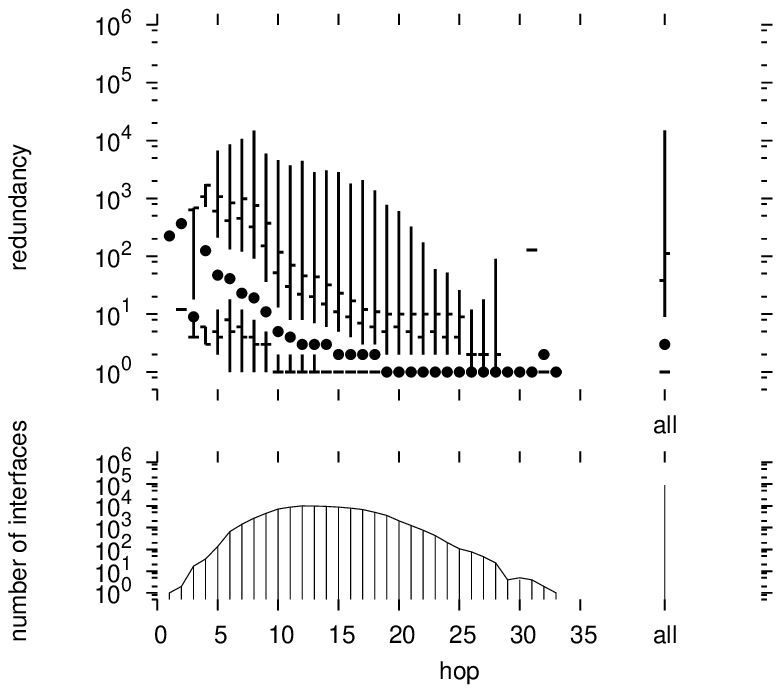}}
    }
    \mbox{
        \subfigure[\texttt{yto} classic]{\label{appendix.redundancy.intra.yto.classic}
            \includegraphics[width=5cm]{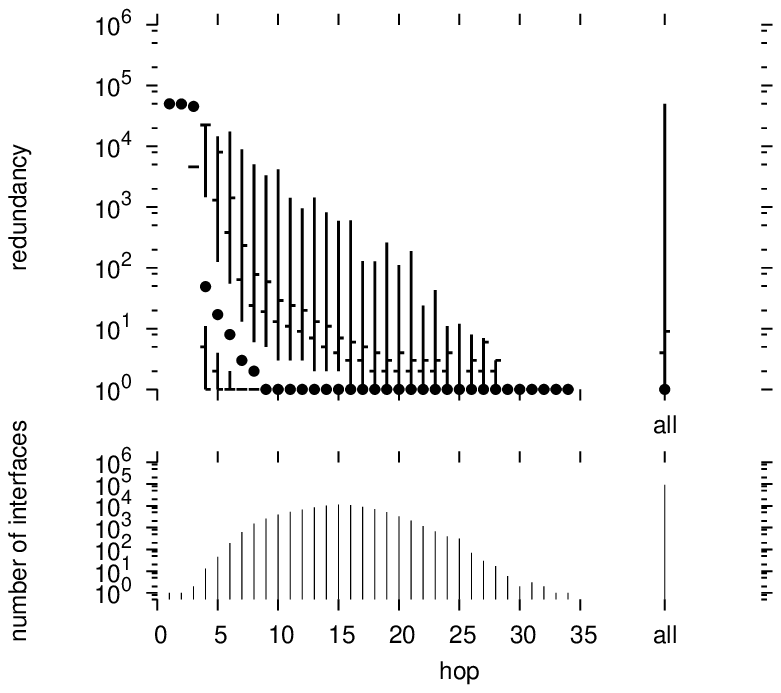}}
            \qquad
        \subfigure[\texttt{yto} Doubletree ($p=0.05$)]{\label{appendix.algo.intra.yto.doubletree.p005}
            \includegraphics[width=5cm]{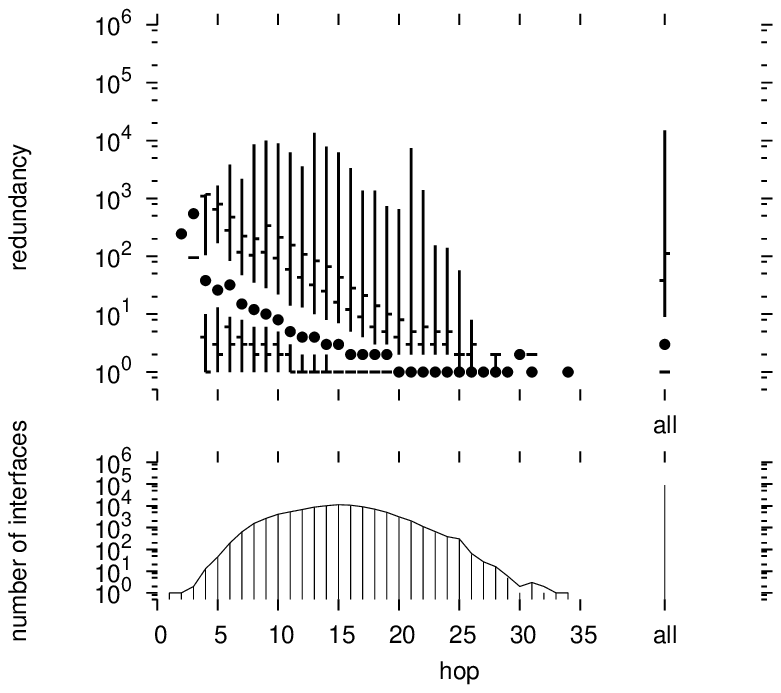}}
    }
    \caption{\texttt{riesling}, \texttt{sjc}, and \texttt{yto}}
  \end{center}
\end{figure*}

\end{document}